\documentclass[preprint2]{emulateapj}

\usepackage{amssymb}
\usepackage{subfigure}
\usepackage{longtable}
\usepackage{epstopdf}
\usepackage{epsfig}
\usepackage{epsf}
\usepackage{amsmath}
\def\lapprox{\mathrel{\hbox{\rlap{\hbox{\lower4pt\hbox{$\sim$}}}\hbox{$<$}}}}
\def\gapprox{\mathrel{\hbox{\rlap{\hbox{\lower4pt\hbox{$\sim$}}}\hbox{$>$}}}}
\def\lesssim{\lapprox}
\def\gtrsim{\gapprox}

\newcommand{\be}{\begin{eqnarray}}
\newcommand{\ee}{\end{eqnarray}}

\newfont{\myfont}{cmmib10}

\newcommand{\btheta}{\hbox{\myfont \symbol{18} }}

\newfont{\myfonttwo}{cmmib8}

\def\arcmin{$^\prime$}
\def\arcsec{$^{\prime\prime}$}
\def\degree{$^{\circ}$}

\newcommand{\cen}{Centaurus~A}

\shorttitle{Faraday rotation structure in the lobes of Centaurus A}
\shortauthors{Feain et al.}

\begin{document}

\title{Faraday rotation structure on kiloparsec scales in the radio lobes of Centaurus A}


\author{I. J. Feain$^{1}$}\email{ilana.feain@csiro.au}
\author{R. D. Ekers$^{1}$}
\author{T. Murphy$^{2,3}$}
\author{B. M. Gaensler$^2$}
\author{J-P Macquart$^4$}
\author{R. P. Norris$^{1}$}
\author{T. J. Cornwell$^{1}$}
\author{M. Johnston-Hollitt$^{5}$}
\author{J. Ott$^{6}$}
\author{E. Middelberg$^{7}$}

\affil{1. CSIRO Australia Telescope National Facility, PO Box 76, Epping NSW 1710, Australia; ilana.feain@csiro.au}
\affil{2. Sydney Institute for Astronomy, School of Physics, The University of Sydney, NSW 2006, Australia}
\affil{3. School of Information Technologies, The University of Sydney, NSW 2006, Australia}
\affil{4. Curtin Institute of Radio Astronomy, Curtin University of Technology, GPO Box U1987, WA 6845, Australia}
\affil{5. School of Chemical and Physical Sciences, Victoria University of Wellington, PO Box 600, Wellington, New Zealand}
\affil{6. National Radio Astronomical Observatory, Charlottesville, P.O. Box O, 1003 Lopezville Road, Socorro, NM 87801-0387 , USA  }
\affil{7. Astronomisches Institut der Universit\"at Bochum, Universit\"atsstr. 150, 44801 Bochum, Germany}


\begin{abstract}
We present the results of an Australia Telescope Compact Array 1.4\,GHz spectropolarimetric aperture synthesis survey of 34 square degrees centred on Centaurus A---NGC~5128. A catalogue of 1005 extragalactic compact radio sources in the field to a continuum flux density of 3\,mJy~beam$^{-1}$ is provided along with a table of Faraday rotation measures (RMs) and linear polarised intensities for the 28\% of sources with high signal-to-noise in linear polarisation. We use the ensemble of 281 background polarised sources as line-of-sight probes of the structure of the giant radio lobes of Centaurus A. This is the first time such a method has been applied to radio galaxy lobes and we explain how it differs from the conventional methods that are often complicated by depth and beam depolarisation effects. Assuming a magnetic field strength in the lobes of $1.3\,B_{1}\,\mu$G, where $B_{1}=1$ is implied by equipartition between magnetic fields and relativistic particles, the upper limit we derive on the maximum possible difference between the average RM of 121 sources behind Centaurus A and the average RM of the 160 sources along sightlines outside Centaurus A implies an upper limit on the volume-averaged thermal plasma density in the giant radio lobes of $\langle n_e \rangle < 5\times10^{-5}B^{-1}_{1}$ cm$^{-3}$ . We use an RM structure function analysis and report the detection of a turbulent RM signal, with rms $\sigma_{\textrm{RM}} =17$\,rad~m$^{-2}$ and scale size $0.3$\degree, associated with the southern giant lobe. We cannot verify whether this signal arises from turbulent structure throughout the lobe or only in a thin skin (or sheath) around the edge, although we favour the latter. The RM signal is modelled as possibly arising from a thin skin with a thermal plasma density equivalent to the Centaurus intragroup medium density and a coherent magnetic field that reverses its sign on a spatial scale of 20\,kpc. For a thermal density of $n_{1}\,10^{-3}$ cm$^{-3}$, the skin magnetic field strength is $0.8\,n^{-1}_{1}\,\mu$G.
\end{abstract}




\section{Introduction}
\label{intro}

The lobes of radio galaxies are magnetised, quasi-freely expanding rarified cavities inflated by relativistic jets propagating outwards through the intergalactic medium from a central supermassive black hole (e.g. \citealp{beg84}). As such, radio lobes could be excellent sites for high-energy particle acceleration and even the production of ultra-high-energy cosmic rays \citep{ben08,fra08,hard09}. Knowledge of the physical conditions in radio lobes, including both the lobe magnetic field strength and thermal plasma density, are important to explore any high-energy acceleration mechanisms in full \citep{kron04b}. Understanding the magnetic and thermal properties of radio lobes in detail is also fundamental to our understanding of galaxy formation in terms of feedback processes between the AGN and the interstellar/intergalactic medium \citep{croton06,cro06,elbaz09}.\newline

Linearly polarised electromagnetic radiation passing through a magnetised thermal plasma causes rotation in the angle of polarisation of the radiation at a rate given by, 
\begin{equation}\begin{split}
\delta\theta &= 0.81~n_e~B_{\parallel}~\delta l~\lambda^2\\
                                               &= \delta RM~\lambda^2
\label{eqn:rm1}
\end{split}\end{equation}
where $\theta$ (in radians) is the position angle of the radiation at wavelength $\lambda$ (in meters), $n_e$ is the thermal electron density (in cm$^{-3}$), $B_{\parallel} $ is the line of sight component of the magnetic field (in $\mu$G), $l$ is the path length through the rotating material (in pc) and RM is the Faraday rotation measure (in units of rad~m$^{-2}$). \newline

If radio galaxy lobes contain magnetised, thermal plasma they will have an associated intrinsic Faraday depth. Observations of RMs in radio sources have shown that internal Faraday rotation in radio lobes is quite small \citep{kron86,kron04b,sch98,palma00} with estimates of the thermal electron densities of $n_e\lesssim 10^{-6}$ cm$^{-3}$, assuming a thermal plasma is distributed uniformly across the lobes. Such low inferred thermal matter densities are orders of magnitude lower than the upper limits on hot (keV) gas obtained from measurements of X-ray cavities around radio lobes \citep{blanton01,fabian00,nulsen05}.  \newline

Whereas only upper limits exist on the uniform thermal matter density inside radio lobes, there is conflicting evidence regarding the presence of a Faraday rotating thin skin (or sheath) around the lobes caused by entrainment of intergalactic plasma. For example, in the case of Cygnus~A, \citet{dreh87} attribute observed RM variations of thousands of rad~m$^{-2}$ wholly to the foreground intracluster medium that Cygnus~A is embedded within. \citet{bicknel90}, however, use the same data to show that this RM structure could arise from a thin skin around the radio lobes where Kevin-Helmholtz instabilities have caused a mixing of the lobe plasma with the intergalactic medium. More recently, a similar debate has arisen as to the origin (intracluster medium versus thin skin) of RMs (and RM variations) in excess of $\pm$1000~rad~m$^{-2}$ across the radio galaxy PKS\,1246$-$410 in the centre of the Centaurus cluster\footnote{The Centaurus cluster at $z=0.01$ is located behind the Centaurus group at $z=0.0018$ that contains the radio galaxy Centaurus A.} \citep{tay02,rud03,ens03,tay07a}. \newline 

There are three distinct scenarios we consider (but see \citealt{burn66}) when using Faraday rotation to probe the properties of a magnetised, thermal plasma:\newline

1. The diffuse polarised synchrotron emission from the lobes is mixed with the magnetised, thermal (Faraday rotating) plasma. In this case, the emission from the back of the source will have been rotated more than the emission from the front of the source and so `depth depolarisation' can occur along any line of sight. In addition, `beam depolarisation' can occur due to variations in RM on scales smaller than the observing beam.  In the former situation (i.e. not applicable to beam depolarisation) the concept of Faraday depth is introduced. Accurate determination of Faraday depth is complex, but necessary to extract information on the magnetic field and thermal density within the source \citep{cioffi80}.  Radio galaxy lobes embedded in clusters or groups are often used to probe the cluster/group medium \citep{dreh87,clarke01,eilek02,tay02,laing2008}. Here one must first show that the rotating plasma arises purely from the foreground medium itself rather than the lobes or skin of the radio source \citep{rud03}.\newline

2. The polarised emission is diffuse and located behind the Faraday rotating plasma. This is similar to the above scenario in terms of beam depolarisation, however no depth depolarisation occurs because there is no mixing of the emitting and rotating regions. For example, extended radio galaxies could be used to directly probe Galactic magnetic fields. \newline

3. The polarised emission is unresolved and located behind the magnetised, thermal (Faraday rotating) plasma. In this scenario, all the polarised signal from any element of the background source is rotated by the entire line of sight through the screen. The screen can cause spatial depolarisation due to variations in RM in the screen across the angular size of the source. This probes scale sizes in the screen on scale sizes smaller than the background source size. No depth depolarisation occurs. This technique is often used to investigate the magnetic structure of the Milky Way \citep{jb03,bro07,mao08}, nearby galaxies \citep{han98,gaen05} and galaxy clusters \citep{kim91,hennessy89}.\newline

Note that scenarios 2 and 3 above are not affected by back-versus-front differences such as the Laing-Garrington effect \citep{laing88,gar88}.\newline

The typical angular size subtended by radio galaxy lobes is too small to include a statistically significant number of compact polarised background sources, at least for the source densities reached with current sensitivity (typically mJy~beam$^{-1}$). Hence, up until now, all studies of the Faraday rotation in radio galaxy lobes have been restricted to using emission from the lobes themselves, as in scenario 1 above (recent examples include \citealp{kharb09,laing2008}). The outer lobes of the nearest radio galaxy, Centaurus A, subtend a large enough angular size ($\approx 45$ deg$^2$) that hundreds of polarised sources are detected along sightlines behind them. For the first time, we can investigate the magnetised plasma in radio lobes --- using scenario 3 --- without the complexities added by depth and beam depolarisation effects. This is the basis for the analysis presented in this paper. \newline

We have recently completed a large spectropolarimetric imaging campaign at 1.4\,GHz with the Australia Telescope Compact Array (ATCA) and the Parkes 64\,m radio telescope, to image in full the polarised structure of the nearest radio galaxy, Centaurus A. The full spectropolarimetric images of Centaurus A from ATCA and Parkes data combined will be reported in a subsequent paper (Feain et al. 2010, in preparation). An additional result of the ATCA component of the observations, we have also observed in full polarisation 1005 compact radio sources in the background of Centaurus A: some along lines of sight through the lobes and some along lines of sight beyond (outside) the boundaries of the radio lobes.  In this paper we present a catalogue of these 1005 compact radio sources. RMs and polarised flux densities are presented for a subset (281) of the sources. We investigate the spatial correspondence between the radio lobes of Centaurus A and the both the distributions of RMs and fractional polarisation of the background sources. \newline

This paper is divided into sections in the following way: \S\ref{observations} describes the ATCA radio continuum observations and \S\ref{calibration} outlines the calibration and data processing procedures. In \S\ref{tara-catalogue}, we define the procedure used for source finding, give the format for the source catalogue and provide the URL where the entire catalogue can be accessed. \S\ref{polderivation} describes the Rotation Measure Synthesis technique with which we derived reliable RMs and polarised flux densities for 281 out of the 1005 sources catalogued. In \S\ref{src-counts} we show the RM distribution and briefly compare the total and polarised intensity source counts from our data with total and polarised intensity source counts from the literature. We also compute a lower limit on the very small scale RM fluctuations from the lobes of Centaurus A.  \S\ref{spatial-variation} presents a detailed investigation into the spatial variations in the ensemble of 281 Faraday rotation measures. We fit and subtract out the Milky Way foreground RM component leaving a residual excess RM dispersion on angular scales $\theta\sim0.3$\degree. We model this excess as possibly arising from a thin skin around the southern giant radio lobe. Finally, our concluding remarks are given \S\ref{conclusion}.\newline

We adopt a distance to Centaurus A of 3.8\,Mpc from \citet{rej04}. At 3.8\,Mpc, 1\degree\ corresponds to $\approx$66\,kpc. \newline

\section{Observations}
\label{observations}

The data used in this paper were obtained as part of a larger ($\sim$45 deg$^2$) radio synthesis imaging survey of Centaurus A (NGC 5128), the nearest active galaxy in the Universe. \newline

The ATCA was used in mosaic mode over four epochs between 2006 December and 2008 March to observe a total field of view covering 5\degree$\times\,$9\degree\  (in 406 pointings) and centred on RA(J2000)~$13^{\rm h}25^{\rm m}27.6^{\rm s}$, DEC(J2000)~$-$43\degree01\arcmin09\arcsec. The standard continuum correlator configuration was employed for the observations at all epochs. This correlator configuration (\textsc{full\_128\_2}) was chosen because it allows dual-frequency observations using 2$\times$128\,MHz bandwidth observing windows split into 32 channels (the spectral resolution is 1.77 channels) per window. We centred the two frequency windows on 1344 and 1432\,MHz; used in this way, the two windows overlap by 40\,MHz and the total useable bandwith is 192\,MHz. Observations were carried out with four complementary array configurations, each with a maximum baseline of 750\,m. The ATCA's linearly polarised feeds measure cross-polarisation data which allowed us to derive the four relevant Stokes parameters ($I, Q, U, V$). The temporal variations in the atmospheric phase were tracked with a two minute observation of {\textsc{PKS B1316$-$46}} about once an hour. In addition, the parallactic angle coverage of {\textsc{PKS B1316$-$46}} was sufficient to enable us to solve for the polarisation leakages in each 12\,hr observing block. {\textsc{PKS B1934$-$638}}  was observed once a day and used both to correct for the instrumental bandpass and derive the absolute flux density scale. On average, each of the 406 pointings in the mosaic received 100$-$120 minutes integration. \newline

\section{Data calibration and post-processing}
\label{calibration}
The data were inspected for radio frequency intereference (RFI) and flagged accordingly using Root Mean Square (RMS)-Based flagging in the automated RFI detection algorithm {\textsc{pieflag}} \citep{mid06}. Approximately 5\% and 30\% of our visibilities at 1384 and 1432\,MHz, respectively, were flagged. Each baseline and channel was then visually inspected and, where residual RFI was evident, manually flagged. \textsc{miriad} was used in the standard way to derive and apply the instrumental bandpass, gain, phase and absolute flux density calibration. The typical level of polarisation leakage in an individual pointing was $\sim0.1\%$ at the phase centre, $\sim0.5\%$ at the half-power point and $\sim1.7\%$ at 0.1-power. The off-axis polarisation leakage of the ATCA is reduced substantially by the combination of mosaicing and tracking a source over a wide range of paralactic angles; we measured the leakage across the calibrated final mosaic to be $\sim0.1\%$. After calibration and disregarding the edge channels, there were 24$\times$ 8\,MHz channels between 1296 and 1480\,MHz. Faraday rotation measures were derived across the full 24-channel band. The ionospheric variations in RM above the ATCA over the course of the observations are typically $\leq1$\,rad~m$^{-2}$ \citep{mac05,bil08}. The variations in the ionospheric RM between observations will cause a slight amount of depolarisation, leading to a few percent decrease in the measured polarisation of the sources. \newline 

In this paper, we concentrate on the polarisation properties of the compact extragalactic radio sources in the background to Centaurus A. For the purposes of analysing these background radio sources, the foreground large scale emission from the giant lobes of \cen\ was considered to be contamination. We therefore filtered out this large scale emission from our data by disregarding all visibilities from baselines shorter than 300\,metres (1.4\,k$\lambda$). The final angular resolution for this data (after filtering) is 60\arcsec$\times$33\arcsec, 3\degree\ (as given in Table~\ref{filtered-sens}). For this work, each of the 406 ATCA pointings were considered, calibrated, deconvolved and restored individually (using the same size beam for all pointings) and then stitched together with linear mosaicing. For each individual pointing, a Stokes $I, Q, U$ and $V$ image was produced (the latter to measure the noise properties in each channel) for every one of the 24 independent (8\,MHz wide) frequency channels and, for Stokes $I$ only, a continuum image was also created using multi-frequency synthesis \citep{sau94}. For the latter Stokes $I$ image alone, a single iteration of phase-only self-calibration was applied; no self-calibration was necessary for the Stokes $Q$ or $U$ images. \newline

\begin{figure*}
\centering
\includegraphics[width=19cm]{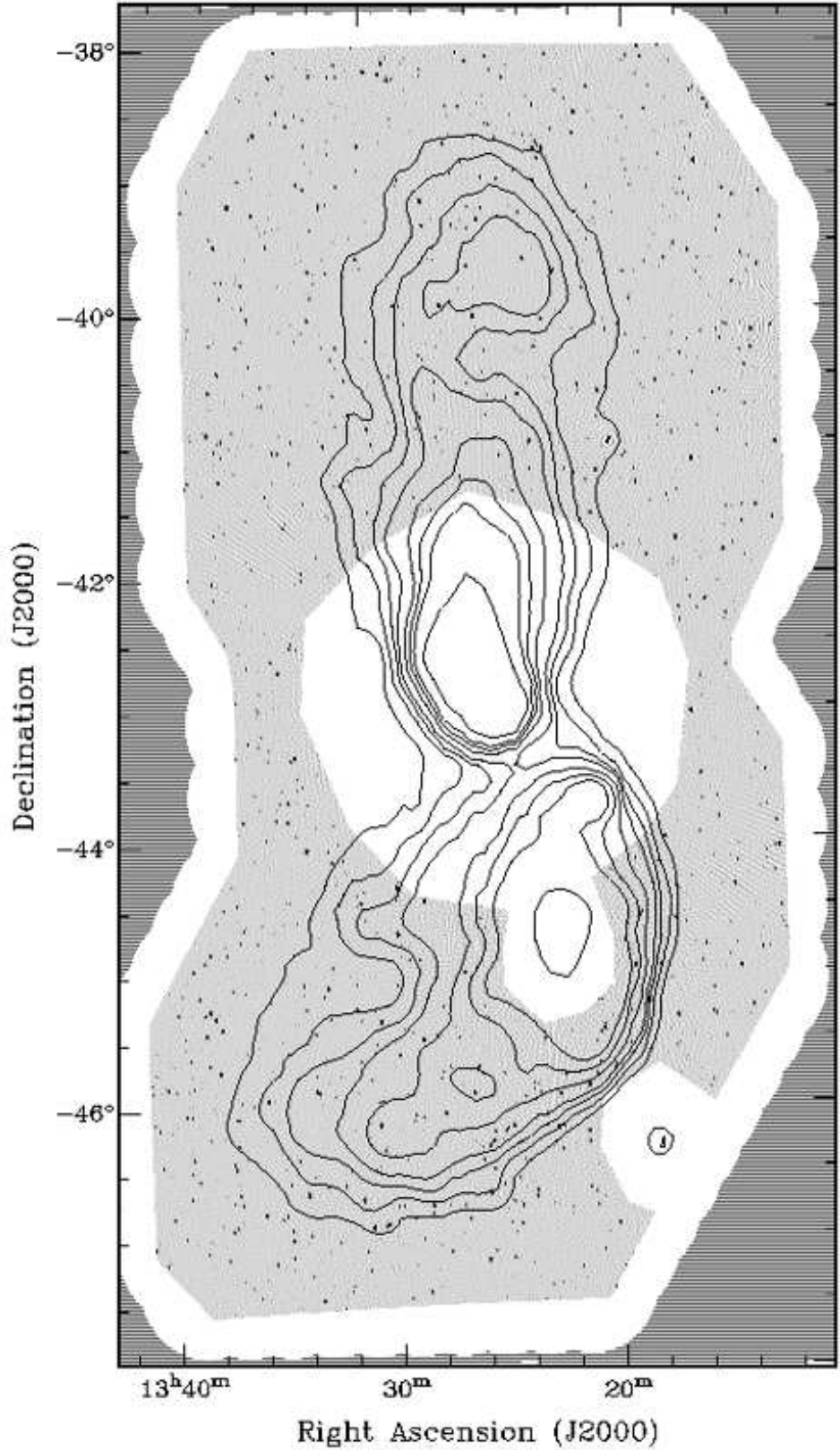}
\caption{The masked, spatially filtered total intensity radio continuum image of the Centuarus A field used to find and catalogue the compact radio sources given in Table~\ref{table:tara-catalogue}. The grey shading shows the regions of our field that we used for source finding. In these regions, many of the compact radio sources can be seen. Masked regions, shown in white, correspond to areas where the sensitivity of the image is poorer, in the vicinity of very bright sources (both the core of Centaurus A as well as two foreground sources with flux densities $>2$Jy  in the southern lobe) or near the edge of the mosaic where primary beam correction was significant. The contours correspond to a Parkes 1.4\,GHz image at 14\arcmin resolution (courtesy Mark Calabretta) with levels $1.5, 2, 2.5, 3, 4, 5, 6, 10, 100$\,Jy~beam$^{-1}$. The horizontal-striped region beyond the edge of the mask that does not contain point sources is beyond the observed mosaic. The image is shown projected in a \textsc{sin} coordinate system.}\label{mask}
\end{figure*}

\begin{figure}

\includegraphics[width=9cm]{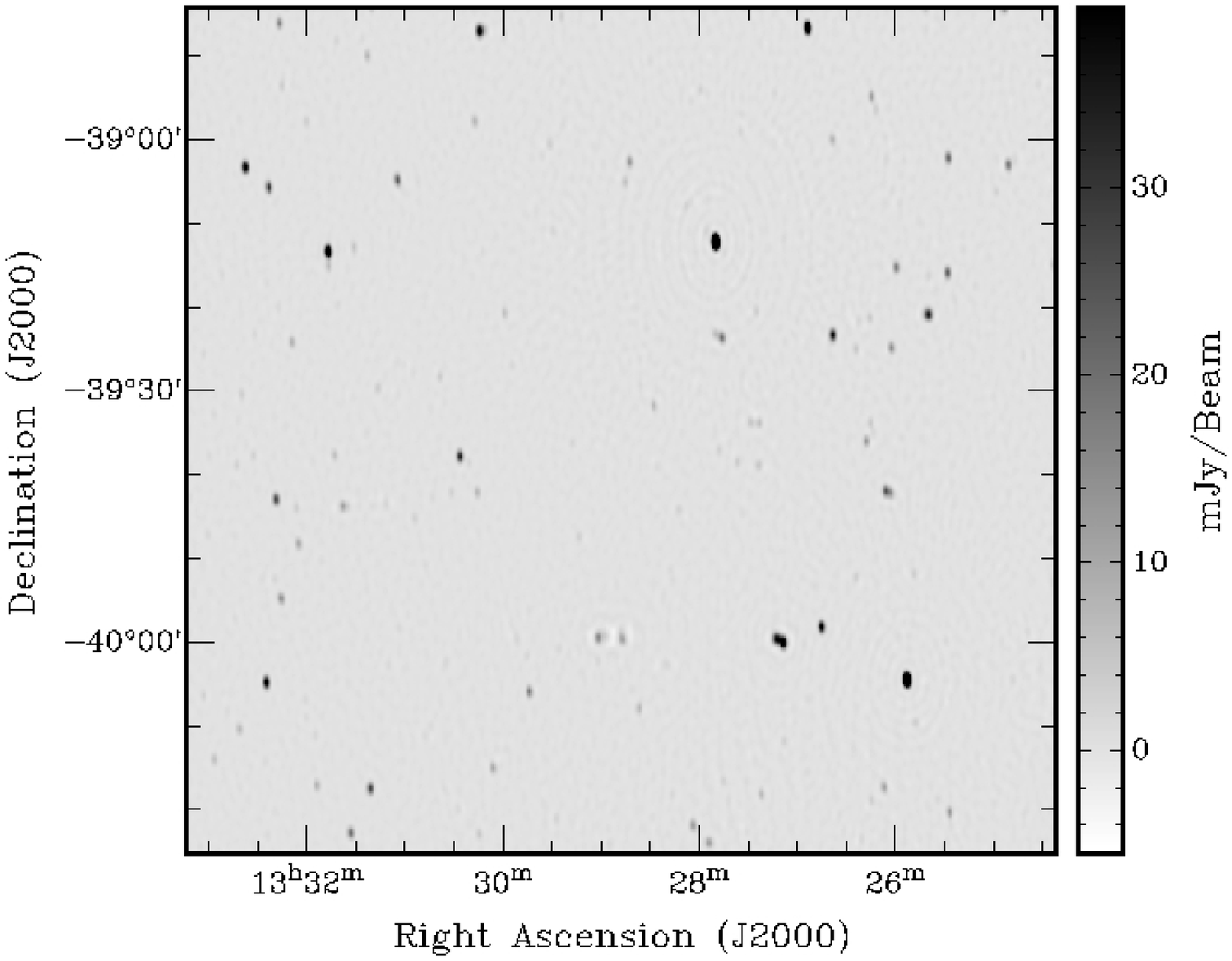}
\caption{A representative portion of the total intensity image shown in Figure~\ref{mask}. The rms level is 0.18\,mJy~beam$^{-1}$.} 
\label{zoom-i}

\includegraphics[width=9cm]{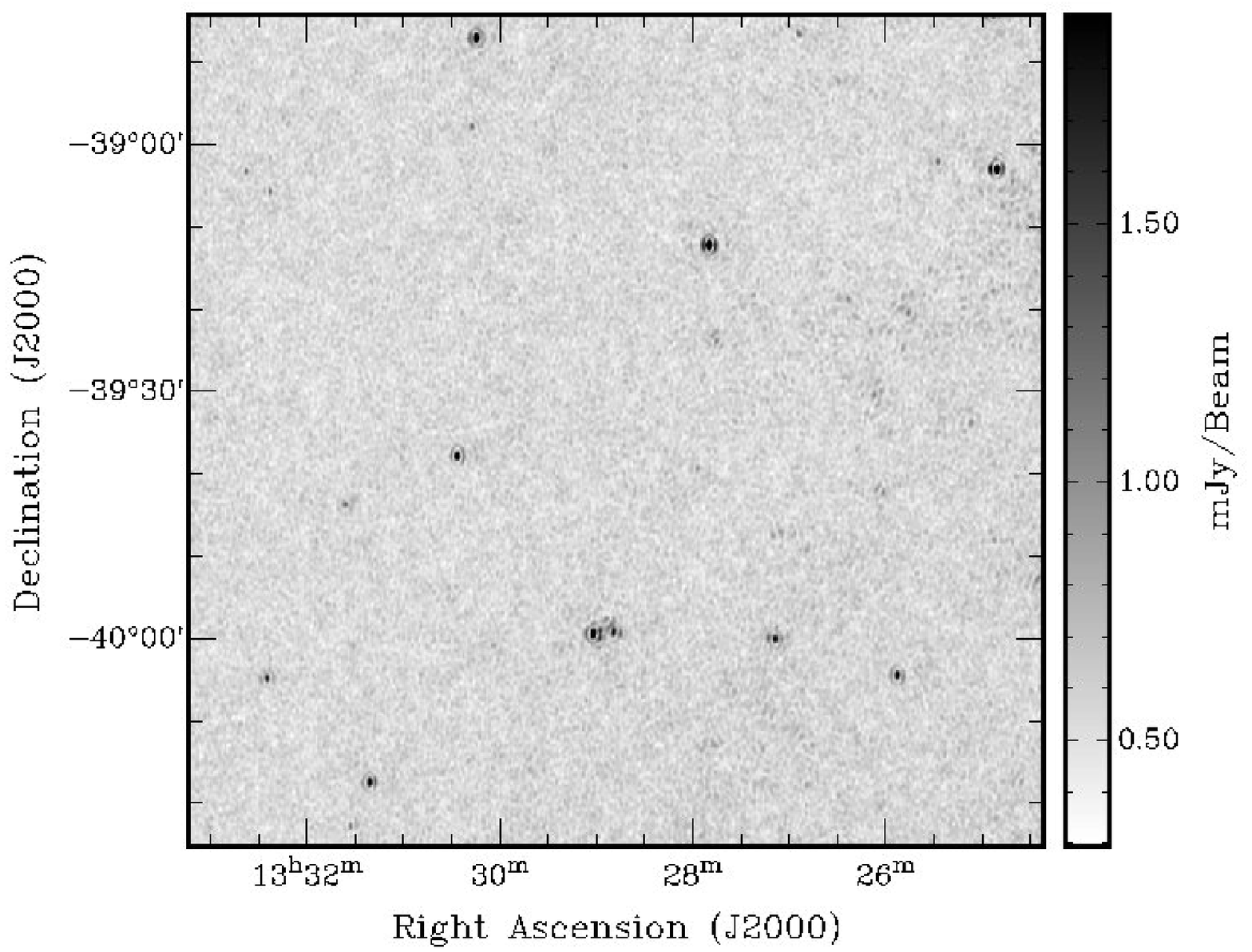}
\caption{As Figure~\ref{zoom-i} but shown in linear polarised intensity, before bias corrrection. The rms level is 0.07\,mJy~beam$^{-1}$}
\label{zoom-p}

\end{figure}

A mask was applied to the spatially filtered dataset in order to create an image with a roughly uniform sensitivity; the point source sensitivities for the final continuum image as well as the individual 8\,MHz Stokes $Q$ and $U$ images are given in Table~\ref{filtered-sens}. Figure~\ref{mask} shows the mask that was applied to the spatially filtered image: the large blanked region in the centre of the mosaic is due to high residual sidelobes around the very bright (S$_{\textrm {peak}} \sim 7.5\,$Jy~beam$^{-1}$ at 1.4\,GHz) central region of Centaurus A. The two smaller blanked regions south of the Centaurus A core are associated with the residual sidelobes of two bright background sources (\textsc{pks~1320$-$446} and \textsc{pmn~j1318$-$4620}). The edge of the mosaic was also masked to remove elevated noise levels ($\gtrsim0.3$\,mJy~beam$^{-1}$) where primary beam correction was significant. Figures~\ref{zoom-i} and \ref{zoom-p} show the continuum intensity and corresponding linear polarised intensity (not corrected for Ricean bias) for a representative portion of the surveyed area. Figure~\ref{mask} has a roughly uniform sensitivity that is given in Table~\ref{filtered-sens}. The total survey area, after masking, is 33.93 deg$^2$. \newline

\begin{table}
\caption{Observational Parameters}
\label{filtered-sens}
\begin{center}
\begin{tabular}{lcc}
\hline
Parameter& Continuum & Stokes ($Q,U$)\\
\hline\hline
Bandwidth (MHz) & 192 & 8 per channel\\
$\sigma$ (mJy~beam$^{-1}$) & 0.15 & 0.5 per channel \\
$\theta_{\textrm {maj}}\times\theta_{\textrm {min}}$ & 63\arcsec$\times$33\arcsec & 63\arcsec$\times$33\arcsec\\
Position Angle & 3\degree & 3\degree\\
\hline
\end{tabular}
\newline \noindent Position angle is defined with north 0\degree\ and east $+90$\degree.
\end{center}
\end{table}

\section{Source Finding}
\label{tara-catalogue}

Source detection was performed using the \textsc{miriad} routine \textsc{sfind} \citep{hop02} on the total intensity image in Figure~\ref{mask}. We ran \textsc{sfind} in its original mode, which uses a `Search and Destory' algorithm much the same as the AIPS task VSAD, to find all sources in the image and fit them with a gaussian. This extracted an initial list of candidate sources with a peak flux density greater than 3\,mJy~beam$^{-1}$ (20$\sigma$). To obtain more accurate fits than those produced by \textsc{sfind} we then ran the \textsc{miriad} task \textsc{imfit} to do a constrained gaussian fit (restricting the fit to a small region around the object) for each source detected in the initial candidate list. We ran this iteratively, subtracting each fitted source from the image, to produce a residual map. We identified poor fits by comparing the rms noise in a small region around the source in the original image and in the residual map. Cases in which the rms noise increased after the fitted source was subtracted from the image were investigated further. Sources with poor fits and sources with multiple components are identified as such in the catalogue. The final catalogue consists of 1005 compact sources to a detection threshold of 3\,mJy~beam$^{-1}$. The associated error in the peak flux density for each source has been estimated by the quadrature sum of the rms noise in the image and the uncertainty in transfering the flux scale of the ATCA primary calibrator \textsc{pks~b1934$-$638} (1$-$2\%\footnote{see ATNF Technical Memo AT/39.3/040.}). The error in the integrated flux density, $\Delta I$, for each source was estimated, using Equation~16 in \citet{con97} to be,
\begin{equation}
\Delta I = \Delta A \times \frac{I}{A},
\end{equation}
where $A$ is the peak flux density. The format of the catalogue in Table~\ref{table:tara-catalogue} is:\\

\noindent \textbf{Column (1):} Source Name.\\
\textbf{Columns (2,3):} Right Ascension and Declination in J2000 coordinates.\\
\textbf{Columns (4,5):} Peak flux density averaged over the full bandwidth in units of mJy and its associated error. \\
\textbf{Columns (6,7):} Integrated flux density averaged over the full bandwidth in units of mJy and its associated error. \\
\textbf{Columns (8,9,10):} Fitted major and minor axis and position angle from \textsc{imfit} \\
\textbf{Columns (11,12,13):} Deconvolved major and minor axis and position angle from \textsc{imfit} \\

\begin{table*}
\caption{Catalogue of the compact continuum radio sources brighter than 3\,mJy~beam$^{-1}$ in the field shown in Figure~\ref{mask}. The table below shows the format of the catalogue and gives the results for the brightest 20 sources; the full catalogue of 1005 radio sources is available electronically.}
\label{table:tara-catalogue}
\tablewidth{0pt}\begin{center}
\begin{tabular}{ccccccccccccc}
\hline\hline
Source &RA (J2000)    &      Dec(J2000)  &    S$_{\textrm {peak}}$ & $\Delta$S$_{\textrm {peak}}$  &S$_{\textrm {int}}$ & $\Delta$S$_{\textrm {int}}$  & B$^{fit}_{maj}$ & B$^{fit}_{min}$ & B$^{fit}_{pa}$ & B$^{decon}_{maj}$ & B$^{decon}_{min}$ & B$^{decon}_{pa}$\\
& hh:mm:ss.ss & ddd:mm:ss.s & mJy\,bm$^{-1}$ & mJy\,bm$^{-1}$ & mJy & mJy & \arcsec & \arcsec & \degree & \arcsec & \arcsec & \degree  \\ 
\hline
131452$-$401056 & 13:14:52.67 & -40:10:56.4 & 708.7 & 7.2 & 761.0 & 7.7 & 63.8 & 35.0 & 3.1 & 11.7 & 9.9 & 84.8 \\
132554$-$464302 & 13:25:54.85 & -46:43:02.3 & 456.3 & 4.6 & 545.2 & 5.5 & 63.4 & 39.2 & 4.1 & 21.3 & 6.4 & 86.3 \\
132340$-$410125 & 13:23:40.83 & -41:01:25.9 & 441.5 & 4.5 & 479.1 & 4.9 & 63.6 & 35.5 & 3.4 & 13.1 & 8.5 & 82.8 \\
133130$-$464508 & 13:31:30.39 & -46:45:08.9 & 430.0 & 4.3 & 514.0 & 5.2 & 63.6 & 39.0 & 4.0 & 21.0 & 8.6 & 85.8 \\
131921$-$443649 & 13:19:21.77 & -44:36:49.5 & 387.1 & 3.9 & 453.3 & 4.6 & 64.1 & 38.0 & 4.1 & 19.2 & 11.2 & 80.2 \\
132748$-$391158 & 13:27:48.95 & -39:11:58.1 & 355.3 & 3.6 & 378.1 & 3.8 & 63.9 & 34.6 & 2.9 & 10.9 & 10.3 & -17.1 \\
132148$-$383251 & 13:21:48.82 & -38:32:51.1 & 308.1 & 3.1 & 328.2 & 3.3 & 63.8 & 34.7 & 3.0 & 10.7 & 10.4 & 84.5 \\
131708$-$385451 & 13:17:08.88 & -38:54:51.4 & 302.0 & 3.1 & 319.0 & 3.2 & 63.8 & 34.4 & 3.0 & 10.0 & 9.8 & -12.0 \\
132301$-$384925 & 13:23:01.68 & -38:49:25.2 & 299.2 & 3.6 & 321.7 & 3.9 & 63.8 & 35.0 & 2.0 & 13.1 & 8.4 & -52.4 \\
133118$-$411956 & 13:31:18.68 & -41:19:56.7 & 278.6 & 4.4 & 444.5 & 7.1 & 65.1 & 50.9 & -7.3 & 40.2 & 12.7 & -75.3 \\
132104$-$411451 & 13:21:04.51 & -41:14:51.3 & 274.4 & 2.8 & 317.1 & 3.3 & 66.0 & 36.4 & 0.1 & 22.0 & 12.0 & -29.8 \\
131710$-$430400 & 13:17:10.90 & -43:04:00.5 & 267.5 & 2.7 & 302.5 & 3.1 & 63.9 & 36.8 & 3.0 & 16.2 & 10.9 & -88.0 \\
132031$-$410338 & 13:20:31.43 & -41:03:38.5 & 267.4 & 12.9 & 326.4 & 15.7 & 61.8 & 41.1 & 8.3 & ----- & ----- & ----- \\
133520$-$391148 & 13:35:20.53 & -39:11:48.4 & 253.9 & 2.6 & 271.7 & 2.8 & 63.9 & 34.8 & 3.0 & 11.1 & 10.7 & -80.1 \\
132551$-$400425 & 13:25:51.63 & -40:04:25.7 & 233.0 & 2.4 & 250.9 & 2.5 & 63.7 & 35.1 & 3.3 & 12.2 & 9.4 & 77.4 \\
131749$-$414536 & 13:17:49.75 & -41:45:36.1 & 226.4 & 2.3 & 249.9 & 2.5 & 63.8 & 36.0 & 3.5 & 14.5 & 9.8 & 80.7 \\
131739$-$452221 & 13:17:39.58 & -45:22:21.5 & 216.0 & 2.2 & 249.1 & 2.6 & 63.0 & 38.1 & 4.1 & ----- & ----- & ----- \\
132936$-$442312 & 13:29:36.61 & -44:23:12.1 & 206.7 & 2.1 & 237.6 & 2.4 & 63.6 & 37.6 & 3.9 & 18.2 & 8.2 & 83.8 \\
131713$-$410934 & 13:17:13.41 & -41:09:34.5 & 197.3 & 2.0 & 223.6 & 2.3 & 63.3 & 37.2 & 3.6 & 17.4 & 5.6 & 86.7 \\
133357$-$464138 & 13:33:57.33 & -46:41:38.0 & 192.9 & 2.3 & 328.0 & 3.9 & 88.9 & 39.8 & 10.8 & 63.4 & 19.9 & 17.1 \\

\hline
\end{tabular}
\end{center}
\end{table*}

\section{Polarisation and Faraday Rotation Measure}\label{polderivation}
The polarised fluxes and Faraday rotation measures (RMs) were derived for 281 of the 1005 catalogued sources listed in Table~\ref{table:tara-catalogue} as follows. \newline

At each source position in Table~\ref{table:tara-catalogue}, we extracted the Stokes~$Q$ and $U$ values and their expected RMS error (based on the sensitivity of the map at that position) in each of the 24 independent spectral channels in our data-set. Converting from frequency to $\lambda^2$, where $\lambda$ is the observing wavelength of each channel, we then have a dependence of the complex Stokes vector $(Q, U)$ as a function of $\lambda^2$ at the peak position of each source. For pure Faraday rotation, this vector should have a phase that varies linearly with $\lambda^2$ at a rate equal to the source's RM. We extracted this RM from each data-set via rotation measure synthesis \citep{bd05,hea09}, in which we compute the Fourier transform of the complex Stokes vector to yield the amplitude of the polarized flux, $P$, as a function of Faraday depth, $\phi$.  For a source with a single valued RM, the RM synthesis spectrum will have a single peak whose height is equal to the polarized flux, and width (RM resolution) determined by the wavelength coverage of the observation. The polarisation position angle is given by the Q and U values at the peak polarised emission. This method achieves signal-to-noise corresponding to the full bandwidth of the observation, but with bandwidth smearing effecting only the individual channel width. Any deviations from $\lambda^2$ behaviour resulting from complex Faraday rotation structure over the background source are seen as structure in the rotation measure synthesis spectrum.  \newline

This analysis was applied to all 1005 sources in Table~\ref{table:tara-catalogue}, with the individual $(Q, U)$ measurements weighted by the inverse square of the sensitivity for each spectral channel.  The total bandwidth and spectral resolution of our data mean that the FWHM in Faraday depth for a single RM component is 280~rad~m$^{-2}$, and that we are sensitive to RMs with magnitudes less than $\approx$3500~rad~m$^{-2}$. The resulting Faraday depth functions exhibit spectral sidelobes because of incomplete wavelength coverage (in the same way that an aperture synthesis image shows sidelobes because of incomplete $u-v$ coverage). Since we can compute the RM transfer function for each source (i.e., the Faraday depth spectrum of a source of unit polarized intensity and zero RM), we can deconvolve our data-set using the same iterative \textsc{clean} approach routinely applied to radio interferometric images \citep{hog74}. Specifically, we have implemented \textsc{rmclean}, as described by \citet{hea09}.\newline

\begin{figure*}
\rotatebox{270}{\includegraphics[width=7.8cm]{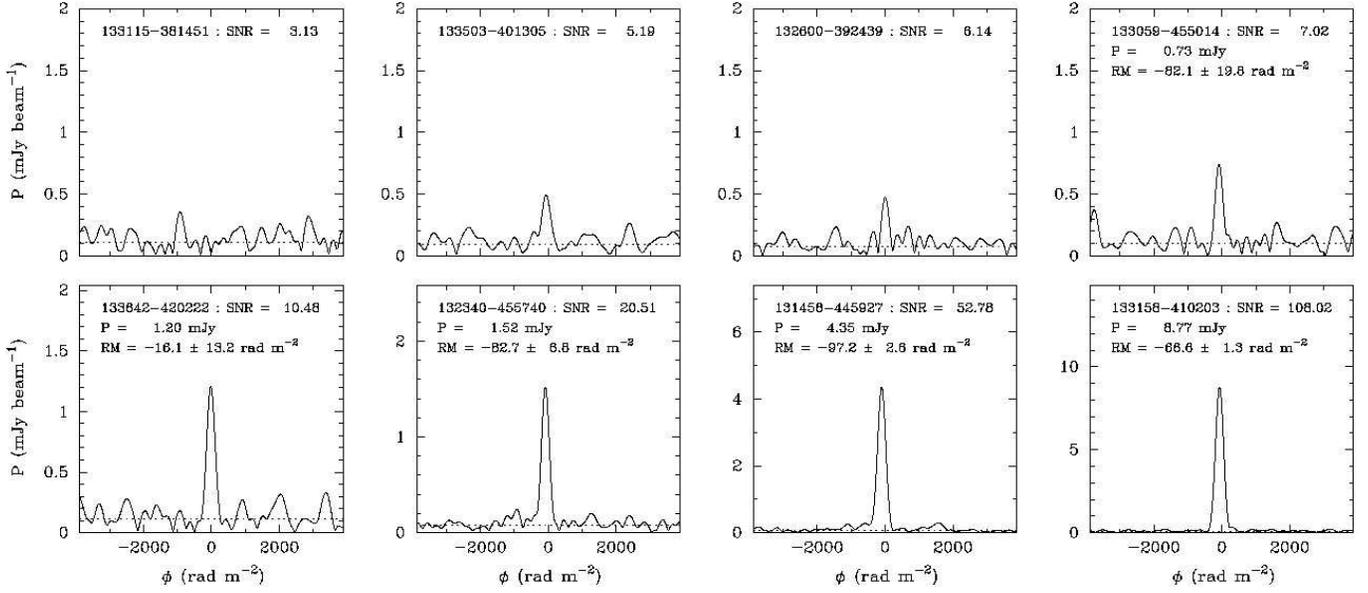}}
\caption{Results of rotation measure synthesis for eight of the 1005 radio sources in Table~\ref{table:tara-catalogue} shown in increasing order of signal-to-noise ratio (SNR) in linear polarised flux density. Each panel plots polarized flux, $P$, vs.\ Faraday depth, $\phi$, at the peak pixel of a source. The solid curve shows the amplitude of the complex polarization vector, while the dotted line shows the 1-$\sigma$ RMS noise level at that position. The legend to each panel provides the source name and the SNR, along with the debiased polarized flux and rotation measure in cases where SNR$\geq7$. }
\label{fig_synth}
\end{figure*}

When the above prescription is applied to all sources, we now have 1005 deconvolved spectra of $P$ (in units of mJy~beam$^{-1}$)\footnote{Strictly speaking, $P$ is in units of surface brightness per unit $\phi$, i.e., mJy~beam$^{-1}$~rad~$^{-1}$~m$^2$, and the polarized flux is determined by integrating $P$ as a function of $\phi$. However, for sources of unresolved Faraday depth the peak surface brightness is equal to the total flux, in the same way that the peak surface brightness of a point source in an image of intensity is equal to its integrated flux.} as a function of $\phi$ (in units of rad~m$^{-2}$). Some examples are shown at varying signal-to-noise in Figure~\ref{fig_synth}.\newline

For each source, we identified the peak value of $P$ as a function of $\phi$, and then applied a parabolic fit around this peak to yield the best-fit estimate of the polarized flux and RM. We then debiased the polarized flux by estimating the observed RMS noise, $\sigma_{QU}$, in the real and imaginary parts of tHe spectrum far from the peak, and subtracting $\sigma_{QU}$ in quadrature from the peak value of $P$ \citep[see][]{ss85}. The ratio of the debiased polarized flux to the noise then yields the signal-to-noise ratio (SNR) of the detection of Faraday rotation.  We adopted a threshold SNR $\ge 7$ as a minimum criterion for a reliable RM determination \citep[see, e.g., Figure~9 of][]{bd05}. For sources with polarized fluxes above this threshold, we computed the uncertainty in RM as the FWHM of the RM transfer function divided by twice the SNR.\newline

Of the 1005 catalogued continuum radio sources, 281 (28\%) polarised sources were robustly detected according to the detection criteria described above and are listed in Table~\ref{poltable}. The format of Table~\ref{poltable} is:\\

\noindent \textbf{Column (1):} Source Name.\\
\textbf{Columns (2,3):} Right Ascension and Declination in J2000 coordinates. \\
\textbf{Columns (4,5):} Peak polarised flux density (after correction for the Ricean bias) in units of mJy\,bm$^{-1}$ and its associated signal to noise ratio (SNR).\\
\textbf{Columns (6,7):} Measured Faraday rotation measure and its associated uncertainty derived using RM synthesis and RM \textsc{clean} in units of rad~m$^{-2}$. \\
\textbf{Columns (8):} Column 6 after correcting for the Galactic contribution (see \S\ref{spatial-variation}) in units of rad~m$^{-2}$. \\


\begin{table*}
\caption{Catalogue of the 281 polarised radio sources detected. The columns in this table are described in \S\ref{polderivation}. The table below gives the brightest 20 sources in linear polarised intensity; the full list of 281 radio sources is available electronically. }
\label{poltable} 
\begin{center}
\begin{tabular}{cccccccr}
\hline\hline
Source & RA (J2000) & DEC (J2000)  & $P$     & SNR &RM& $\Delta$RM$^{*}$ &RM$_{cor}^{**}$\\
 &hh:mm:ss.ss & ddd:mm:ss.s  &  (mJy\,bm$^{-1}$) & &  rad\,m$^{-2}$ & rad\,m$^{-2}$& rad\,m$^{-2}$ \\
\hline
133357$-$464138 & 13:33:57.34 & $-$46:41:38.04 & 13.95 & 154.2 & $-$59.5 & 0.9 & 1.6  \\
131713$-$410934 & 13:17:13.42 & $-$41:09:34.56 & 12.98 & 123.4 & $-$74.6 & 1.1 & $-$19.6\\
132937$-$444416 & 13:29:37.75 & $-$44:44:16.44 & 12.47 & 129.9 & $-$77.7 & 1.1 & $-$20.8\\
133628$-$403529 & 13:36:28.46 & $-$40:35:29.40 & 10.06 & 95.9 & $-$15.2 & 1.4 & 20.6 \\
131749$-$414536 & 13:17:49.75 & $-$41:45:36.00 & 9.43 & 76.0 & $-$30.5 & 1.8 & 25.9 \\
132543$-$461919 & 13:25:43.01 & $-$46:19:19.56 & 9.14 & 122.9 & $-$47.2 & 1.1 & 19.0 \\
133158$-$410203 & 13:31:58.13 & $-$41:02:03.12 & 8.77 & 108.0 & $-$66.6 & 1.3 & $-$25.3\\
133027$-$452406 & 13:30:27.55 & $-$45:24:06.12 & 8.06 & 84.7 & $-$62.3 & 1.6 & $-$3.6 \\
132553$-$462121 & 13:25:53.71 & $-$46:21:21.96 & 7.79 & 94.7 & $-$42.2 & 1.5 & 24.0 \\
133800$-$452717 & 13:38:00.26 & $-$45:27:17.64 & 7.66 & 116.6 & $-$40.7 & 1.2 & 12.3 \\
131921$-$443649 & 13:19:21.77 & $-$44:36:49.68 & 7.52 & 67.4 & $-$73.0 & 2.1 & $-$24.7\\
133241$-$450802 & 13:32:41.71 & $-$45:08:02.04 & 7.52 & 101.9 & $-$80.6 & 1.4 & $-$7.9 \\
132145$-$403618 & 13:21:45.46 & $-$40:36:18.36 & 6.53 & 90.1 & $-$71.2 & 1.5 & $-$22.4\\
133001$-$413824 & 13:30:01.15 & $-$41:38:24.00 & 6.42 & 51.5 & $-$53.2 & 2.7 & $-$8.0 \\
132827$-$464808 & 13:28:27.74 & $-$46:48:08.28 & 6.20 & 89.2 & $-$47.7 & 1.6 & 18.1 \\
132253$-$461330 & 13:22:53.52 & $-$46:13:30.72 & 6.12 & 71.9 & $-$57.7 & 1.9 & 10.4 \\
133746$-$451641 & 13:37:46.54 & $-$45:16:41.52 & 6.05 & 69.6 & $-$56.8 & 2.0 & $-$4.3 \\
131350$-$410134 & 13:13:50.95 & $-$41:01:34.32 & 6.02 & 29.6 & $-$67.0 & 4.7 & $-$9.4 \\
131630$-$444944 & 13:16:30.31 & $-$44:49:44.76 & 5.79 & 72.8 & $-$97.9 & 1.9 & $-$29.5\\
133107$-$465745 & 13:31:07.06 & $-$46:57:45.36 & 5.57 & 69.8 & $-$18.9 & 2.0 & 45.4 \\
\hline                                                                        
\end{tabular}
\\ * The associated uncertainty in RM does not include the uncertainty due to ionospheric variations which are $\leq1$\,rad~m$^{-1}$.
\\ ** RMs after correcting for the Galactic contribution.
\end{center}
\end{table*}

\begin{figure}
\includegraphics[width=8cm]{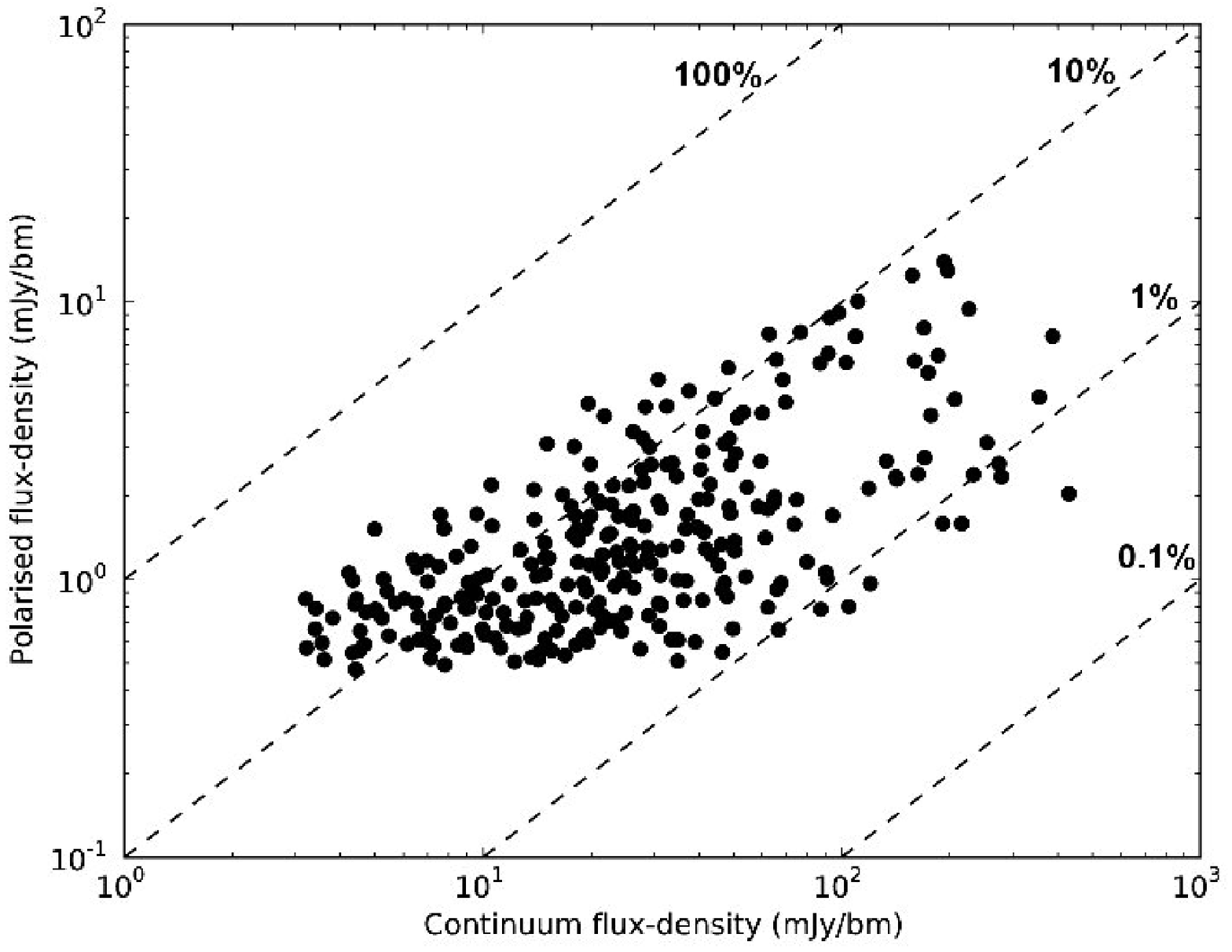}
\caption{Total versus linearly polarised flux density plane for the 281 sources in our sample whose debiased polarised intensities correspond to SNR$\geq$7; see \S\ref{polderivation} for a justification of this threshold. The dashed, diagonal lines represent lines of constant fractional polarisation (from top left to bottom right: 100\%, 10\%, 1\%, 0.1\%).}
\label{ip-plane}
\end{figure}

Figure~\ref{ip-plane} shows the distribution in the total intensity to polarised intensity plane of the 281 sources in our sample whose debiased polarised intensities have a SNR$\geq$7. The dashed, diagonal lines represent lines of constant fractional polarisation. 

\subsection{Which sources are behind the lobes?}\label{inoutdefine}
For much of the rest of this paper, we wish to analyse separately the polarised sources located behind the lobes of Centaurus A from polarised sources whose sightlines pass outside of the lobes. The latter are used as a control sample and to model and subtract the RM component from the Milky Way. We define  what constitutes \textit{behind} the radio lobes to correspond to an `edge' where, at 1.4\,GHz and 14\arcmin\ resolution, the surface brightness of the lobes drops below 1.5\,Jy~beam$^{-1}$; this is the last contour in Figures~\ref{mask} and \ref{rm-distribution}. At approximately this value, the surface brightness of the lobes decreases sharply (but there is clearly still diffuse emission from the lobes present, probably out to the edges of our mosaic). Using a 1.5\,Jy~beam$^{-1}$ threshold, 121 out of the total 281 RMs are behind the radio lobes with the remaining 160 RMs along sightlines outside the lobes. We have tested the sensitivity of our results to the exact value of the boundary chosen for inside versus outside the lobes. We are confident that our result is robust for all contour values that we tested as boundaries (which were between 1 and 2~Jy\,beam$^{-1}$).

\section{RM distribution \& Source Counts}
\label{src-counts}
\subsection{RM Distribution}\label{rmdistributuion}
\begin{figure}
\centering
\includegraphics[height=7cm]{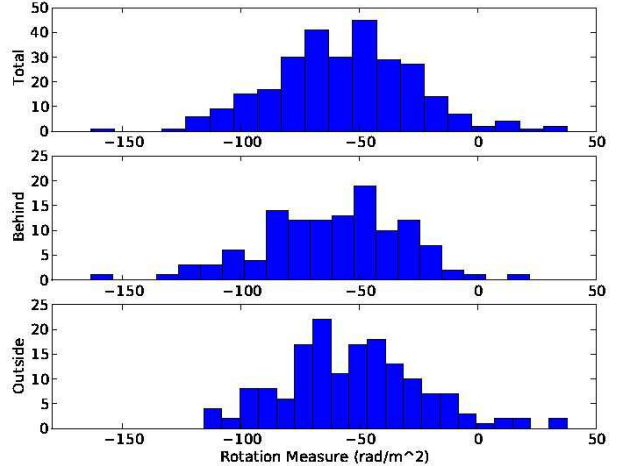}
\caption{\textit{Top panel}: RM distribution of the 281 sources in Table~\ref{poltable}. The mean RM is $-57$ rad\,m$^{-2}$ and the standard deviation is 30~rad\,m$^{-2}$. \textit{Middle panel}: RM distribution of the 121 sources behind the lobes (mean $-61.3$~rad\,m$^{-2}$ and standard deviation  29.8~rad\,m$^{-2}$). \textit{Bottom panel}: RM distribution of the 160 sources outside the lobe (mean $-52.9$~rad\,m$^{-2}$ and standard deviation  29.2~rad\,m$^{-2}$). The bin size is 20~rad\,m$^{-2}$ in all panels.}\label{RM_histo}
\end{figure}

A histogram of the distribution of the full RM sample is shown in the top panel of Figure~\ref{RM_histo}. The mean RM of the total distribution is $-57$\,rad~m$^{-2}$ with a standard deviation of $30$\,rad~m$^{-2}$. These values are in good agreement with earlier observations of the Faraday rotation measure of the lobe emission of Centaurus A reported to vary from $-40$ to $-70$ rad~m$^{-2}$ across the source \citep{cpc65,clarke92}. Other studies have also reported an average RM for Centaurus A of $-59\pm3$rad~m$^{-2}$ \citep{gd66,snkb81}. The RM of Centaurus A is typical of that measured for other nearby sources which implies that the RM is dominated by a foreground component \citep{mjh04, short07}. The middle and bottom panels of Figure~\ref{RM_histo} show histograms of the RM distributions of sources behind and outside the lobes of Centaurus A, as defined in \S\ref{inoutdefine}. \newline

Examination of the available RM sky data from interpolated maps \citep{mjh04, Frick01} shows the that Centaurus A is embedded in a large region of negative rotation measure of average value $\sim$ -27 rad $m^{-2}$ with a slope of roughly -0.9 rad $m^{-2}$ per degree in longitude and -0.6 rad $m^{-2}$ per degree in latitude over the region corresponding to these observations. Unfortunately, the paucity of rotation measure data in the southern hemisphere means interpolated values in this region have been inferred from some of the the most sparse data distributions in the entire sky. 

\subsection{Source counts}

We have compared our total and linearly polarised intensity source counts as a function of flux to published source counts in order to confirm that our sample is statistically comparable to other samples in the literature and that Centaurus A in the foreground is not affecting the global statistics of the sample. We find:
\begin{enumerate}
\item{Our derived total intensity source counts are consistent, within the errors, with the continuum source counts from the NRAO VLA Sky Survey \citep{con98}.}
\item{Our polarised intensity source counts (in the flux range $0.5\leq P\leq 14$\,mJy) are consistent within the errors with the polarised source counts from the ELAIS N1 field (in the flux range $0.5\leq P\leq 20$\,mJy) reported by \citet{tay07b}. }
\end{enumerate}

\subsection{A limit on small scale turbulence in the lobes}\label{invsout}
Random fluctuations in either $B_{\parallel}$ or in $n_e$ on some angular scale R will increase the dispersion in RM determined on this same scale. If R is smaller than the resolution element used to probe the fluctuations, then this effect will lead to `beam' depolarisation where the `beam' size is the angular scale of the background polarised sources. Such small scale RM fluctuations have been inferred in the Galactic plane \citep{haverkorn08} and the Large Magellanic Cloud \citep{gaen05} using an ensemble of extragalactic polarised sources.  \newline

We find no significant difference in the polarised source counts for the 121 sources behind the radio lobes of Centaurus A compared with the 160 sources outside the lobes. The lack of detection corresponds to a limit on magnetised thermal structure in the lobes of Centaurus A on scales much smaller than the median angular scale size of the background sources themselves; this is $\sim10$\arcsec\ for a 10\,mJy source at 1.4\,GHz \citep{win84}. The mean fractional polarisation of the 121 polarised\footnote{Note that we analyse those polarised sources with a SNR$\geq$7 only; there are polarised sources below this SNR cutoff but we do not consider them in this study.} sources behind the lobes (quoted with the standard error in the mean) is $7.6\pm0.5$\% compared to the mean fractional polarisation for the sources outside the lobes of $7.3\pm0.4$\%.  Using the \citet{burn66} law for depolarisation, 
\begin{equation}
f_d = e^{-2\sigma^2_{RM}\,\lambda^4}, 
\end{equation}
and a conservative limit of $f_d>0.7$ on the ratio of the observed to intrinsic polarisation of sources behind the lobes, we derive an upper limit on the very small scale RM turbulence in the giant radio lobes of Centaurus A of $<10$\,rad\,m$^{-2}$ on scales $\ll$180\,pc. A similar limit holds for the volume-averaged RM amplitude in the radio lobes (\S\ref{ordered}), implying that within the limits of our data the ordered component of the lobes has no prefered scale size.


\section{Spatial Variations in the Rotation Measure}
\label{spatial-variation}

\begin{figure*}
\centering
\includegraphics[width=19cm]{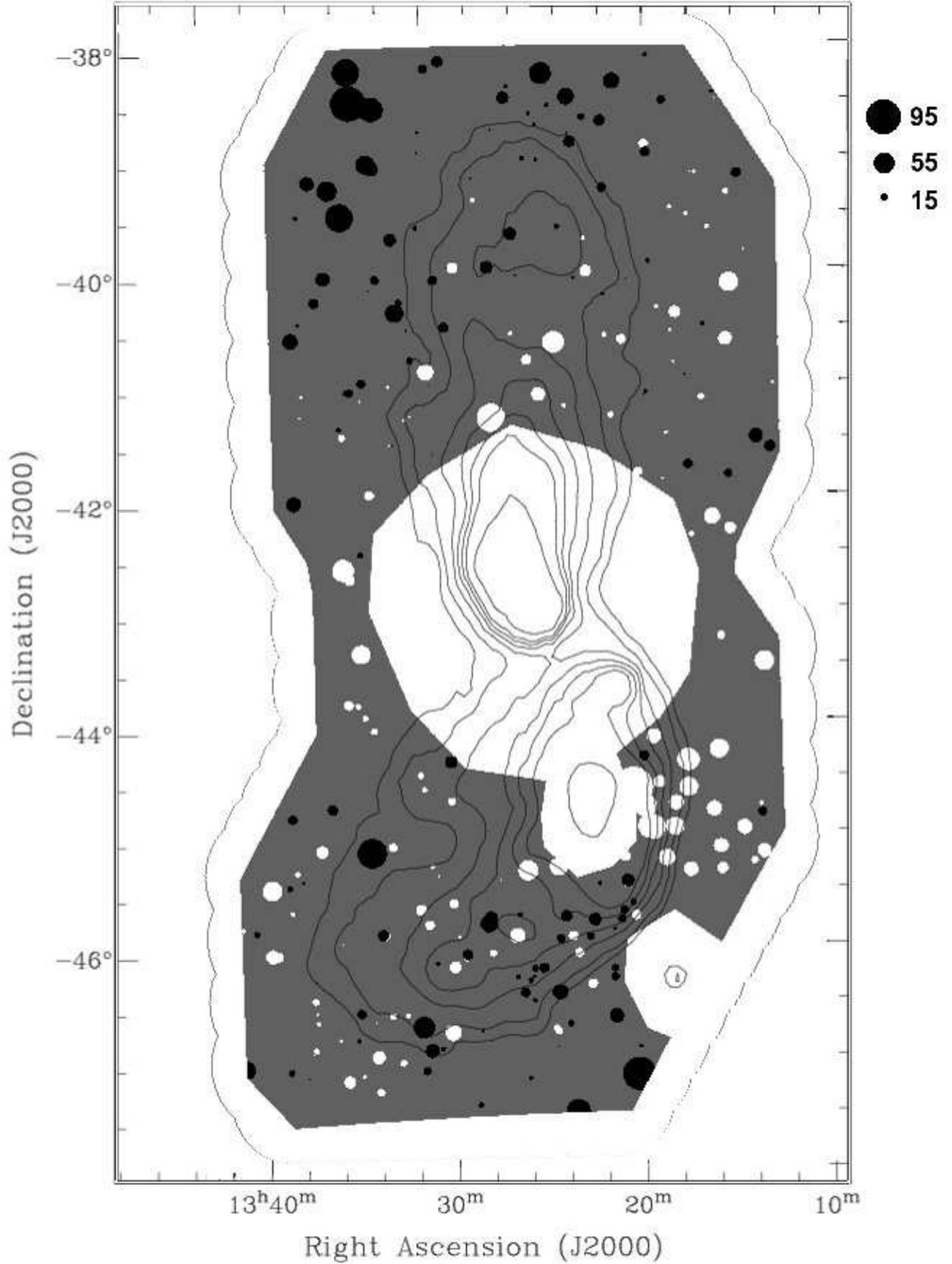}
\caption{Locations and RMs of the 281 sources in Table~\ref{poltable}. To better highlight the variations, the diameter of the sources represent the amplitude of their residual RM after the mean RM of the whole distribution ($-57$\,rad~m$^{-2}$) has been subtracted. Black and white sources are those with positive and negative residuals from the mean, respectively. Overlaid are Parkes 1.4\,GHz radio continuum contours of Centaurus A. Contour levels are 1.5, 2, 3, 4, 5, 6, 10, 100\,Jy\,beam$^{-1}$. The legend on the right hand side of the figure shows the relation between the source diameter and the absolute value of the mean-subtracted RM in units of \,rad~m$^{-2}$.}
\label{rm-distribution}
\end{figure*}

The spatial distribution of background RMs is represented in Figure~\ref{rm-distribution}. The size of the disks correspond to the value of the RM relative to the mean of the sample ($-57$\,rad~m$^{-2}$). Black and white sources are those with positive and negative residuals from the mean, respectively. \newline

A non-zero RM component associated with the lobes of Centaurus A could result from thermal material, mixed with non-thermal material, in the lobes or, perhaps, from the thin skins of ionised plasma around the edges of the radio lobes that are created when the radio lobes inflate cocoons, sweeping up the thermal plasma as they expand into the intergalactic medium. There will be an excess RM if the plasma retains an ordered component and an excess RM dispersion in the presence of a turbulent component.\newline

In this section, we are interested in separating the Milky Way RM component from any RM component that is intrinsic to the radio lobes of Centaurus A. Although it is not possible to do this for the RM of an individual source, it is possible to isolate the Milky Way and Centaurus A RM contributions in a statistical sense, subject to the assumption that the statistics of the Milky Way spatial RM fluctuations on small scales (less than a few degrees) do not change appreciably across the field. To do this we use the 160 RMs that are not located along sightlines through the radio lobes (\S\ref{inoutdefine}) to fit for the foreground Milky Way component. This was achieved using the \textsc{miriad} task \textsc{impoly} to fit a first order polynomial to the RM surface in two dimensions. This surface was then subtracted from all 281 RMs. In this way, we have essentially also removed the Milky Way RM component from the RMs located behind the radio lobes of Centaurus A. 


\subsection{The ordered component}\label{ordered}
The measured RMs of the 121 sources behind the lobes of Centaurus A have a mean of $-61.3$~rad\,m$^{-2}$ and a standard deviation of 29.8~rad\,m$^{-2}$. After removing the Milky Way RM component using the procedure described above, the mean (quoted with the standard error in the mean) of the sources reduces to $-4.5\pm2.8$~rad\,m$^{-2}$ and the standard deviation increases slightly to 30.4~rad\,m$^{-2}$.  The measured RMs of the 160 sources located along sightlines outside the lobes have a mean of $-52.9$~rad\,m$^{-2}$ and a standard deviation of 29.2~rad\,m$^{-2}$. Following subtraction of the Milky Way RM component, the mean reduces to $-1.6\pm2.1$~rad~m$^{-2}$ and the standard deviation decreases to 26.6~rad\,m$^{-2}$. The $3\sigma$ upper limit on the difference between the mean RM inside and outside the lobes ($\sim10$\,rad\,m$^{-2}$) is consistent with the limit infered for very small scale RM structure in the lobes (\S\ref{invsout}), implying that within the limits of our data the ordered component of the lobes has no prefered scale size. To estimate the uniform thermal plasma density in the lobes, we assume that the lobe magnetic field is dominated by an uniform component with strength $B_{\parallel}=1.3\,B_{1}\,\mu$G (setting $B_{1}=1$ gives the equipartition field strength; \citealp{hard09}) and N possible reversals of the field, the path length through the lobes to be $D=200$~kpc, and the upper limit on the volume-averaged RM amplitude associated with the lobes to be 10~rad\,m$^{-2}$. Substituting these values into Equation~\ref{eqn:rm1}, we derive an upper limit on the uniform thermal plasma density in the lobes of $\langle n_e \rangle < 5\times10^{-5}\,B^{-1}_{1}\,\sqrt{N}$ cm$^{-3}$. For $N=1$, this limit is $2-3$ times smaller than previously published limits \citep{mar81,iso01} which in turn further constrains the expected $\gamma$-ray emission in the lobes and the prospects for the production of ultra-high energy cosmic rays \citep{hard09}.

\subsection{The turbulent component}
The difference between the standard deviation of the foreground-subtracted RMs behind the lobes compared to outside the lobes ($3.8$~rad~m$^{-2}$), if real, implies structure/turbulent magnetised plasma in the lobes. To verify and quantify this difference, we have compared the amplitude of the RM fluctuations as a function of angular separation in the lobes (or around the edges of the lobes) to  a `control group' of RMs located on sightlines outside the lobes using an approach based on the structure function,
\be
D_{\rm RM}(\btheta) = \langle \left[ {\rm RM}(\btheta + \btheta') - 
{\rm RM}(\btheta') \right]^2 \rangle.
\label{SFeq}
\ee
where $\btheta$ is the angular separation between sources. The structure function is a robust and reliable means of measuring rotation measure fluctuations \citep{simon84}. In comparison with other statistical techniques like an autocorrelation function analysis for example, the structure functions is far less susceptible to uncertainties in the mean RM level and to large-scale RM gradients, on scales below those comparable to the size of the gradient. A structure function-based analysis is also immune to the irregular spatial sampling of our RMs.  \newline

\subsubsection{Rotation measure structure functions}
\label{distinguish}

Prior to forming structure functions for sources behind and outside the lobes, the large-scale RM component from the foreground Milky Way was fit and subtracted, as described above. When we compared the structure function formed from the foreground-corrected RMs to the structure function formed from the observed RMs (i.e. with no foreground subtraction applied), we found no significant difference between the two results for scales below $\approx4$\degree.  This is unsurprising because the large scale gradients are expected to affect the structure function on scales comparable to the scale of the gradient only.  \newline

\begin{figure*}
\centering
\mbox{\subfigure[]{\includegraphics[width=5.8cm]{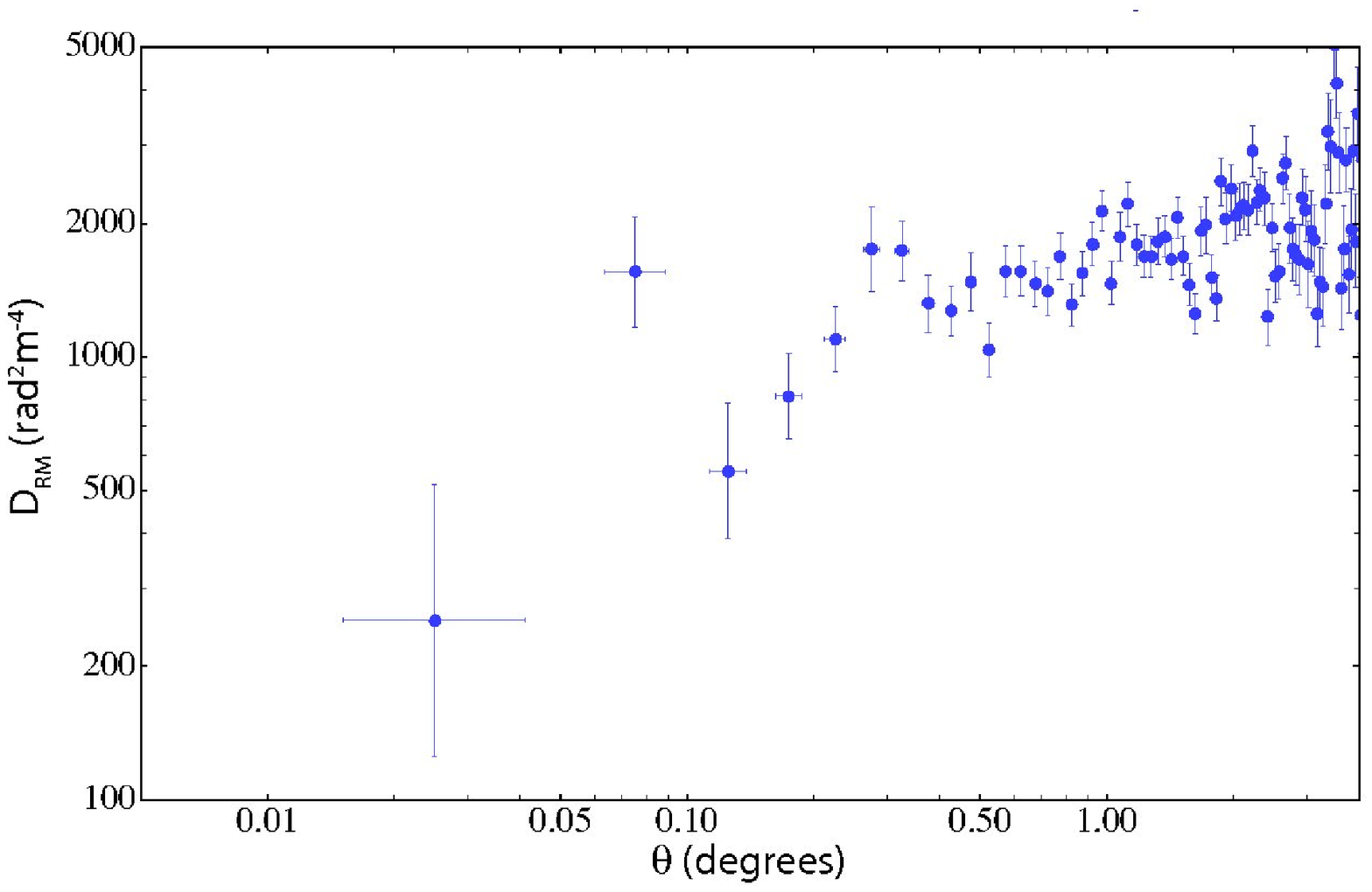}}\quad
\subfigure[]{\includegraphics[width=5.8cm]{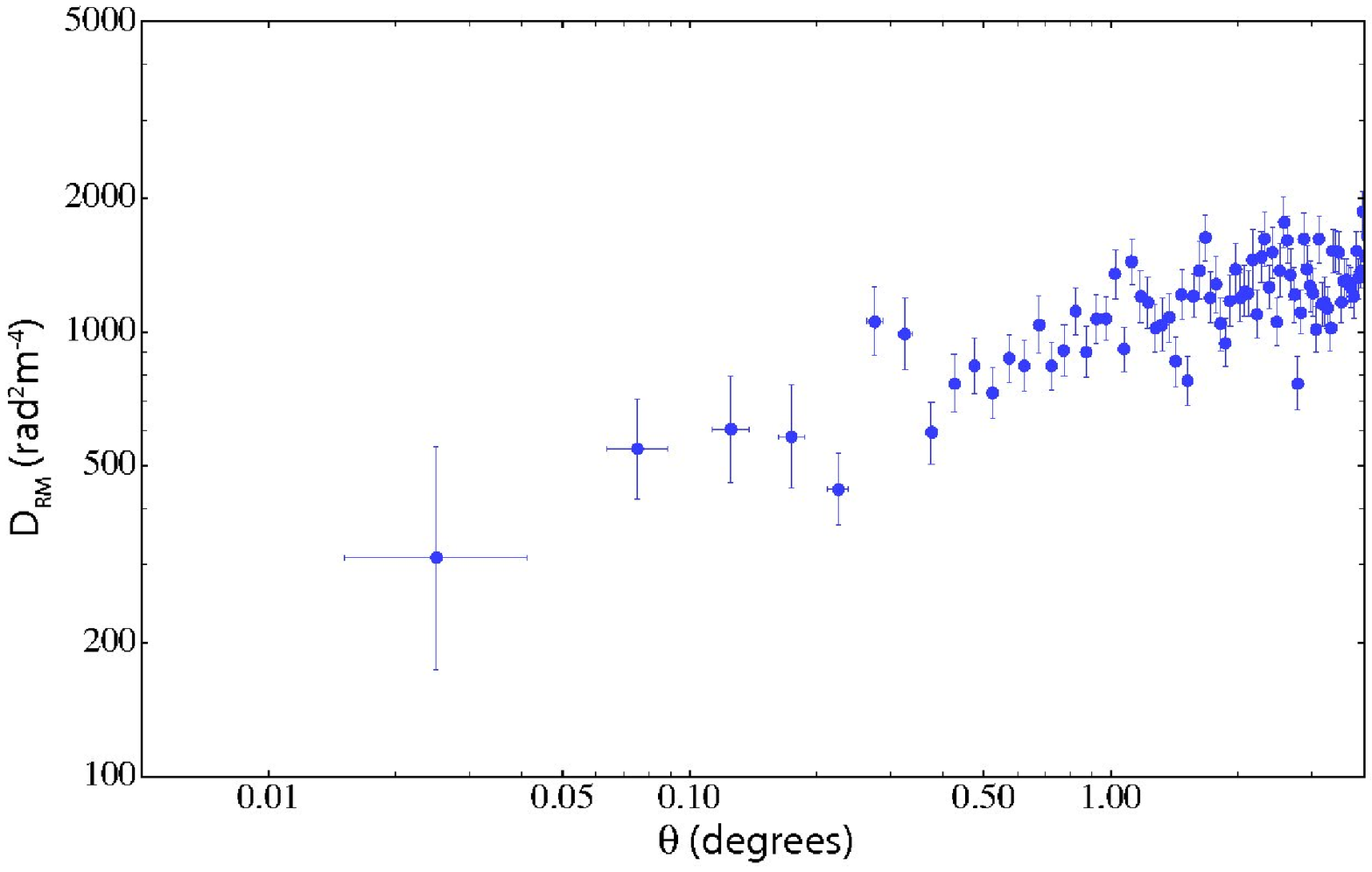}}\quad
\subfigure[]{\includegraphics[width=5.8cm]{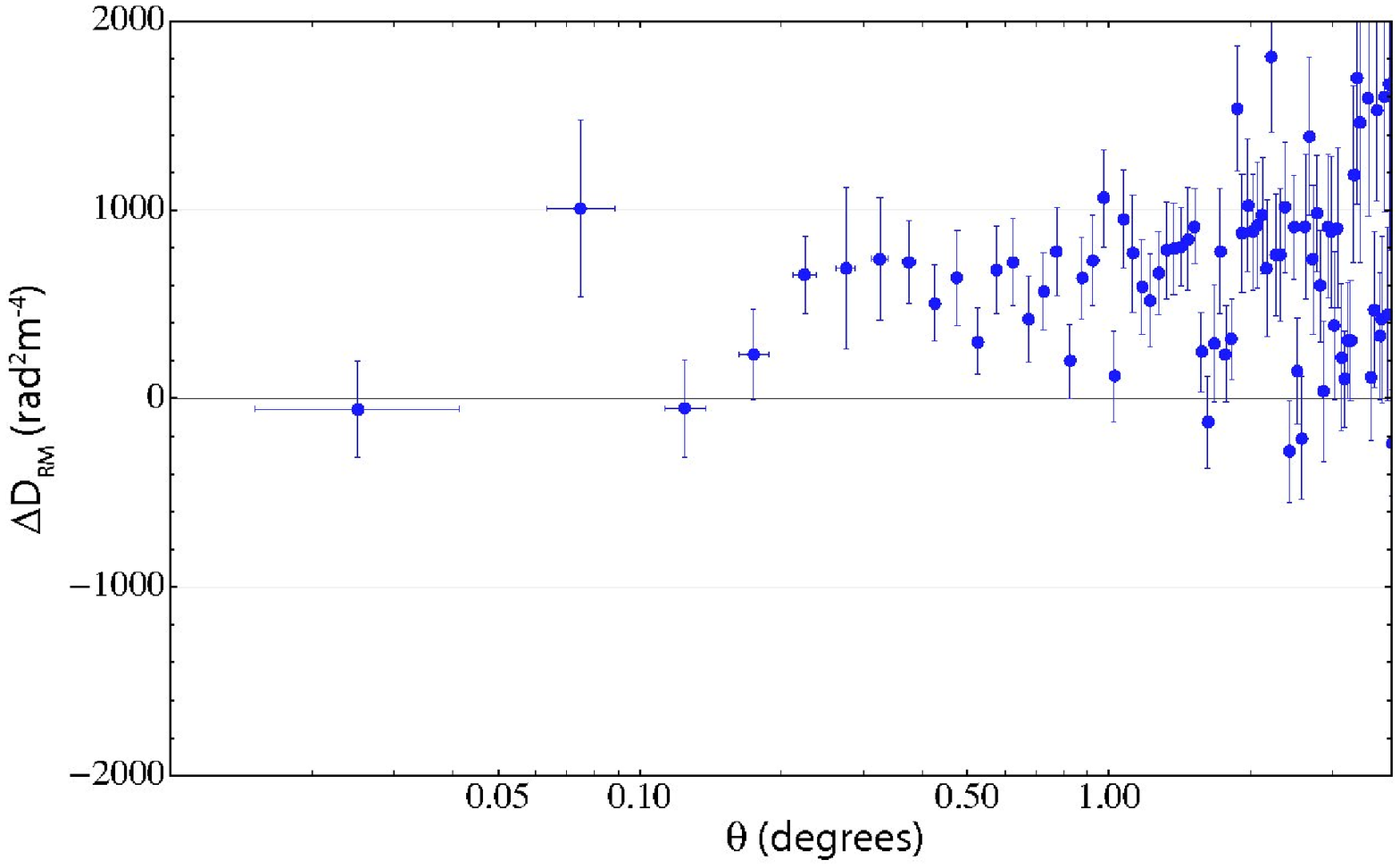}}}
\caption{The RM structure functions of (a) the 121 polarised sources located behind the radio lobes of Centaurus A and (b) the 160 polarised sources located along sightlines outside the radio lobes of Centaurus A. (c) The difference between (a) and (b).}
\label{InOutDiffLobe}
\end{figure*}

Rotation measure differences, as probed by a structure function, can arise from intrinsic differences in the radio sources themselves, or due to spatial variations in the foreground (i.e. in the Galaxy or in the lobes of Centaurus A). Figures~\ref{InOutDiffLobe}a and \ref{InOutDiffLobe}b are the rotation measure structure functions for sources located behind and outside the radio lobes of Centaurus A, respectively.  It is clear from a comparison of Figures~\ref{InOutDiffLobe}a and \ref{InOutDiffLobe}b that the RMs behind the lobes have excess structure on scales $ 0.2^\circ \lesssim \theta \lesssim 4.0^\circ$. The point at the smallest angular scale, $\theta=0.025^\circ$,  is determined solely on the basis of rotation measure differences between each of the double sources in our sample. Since many of these double sources are likely to be the lobes of an individual galaxy, variatons in RMs between the double sources is likely to be much lower than intrinsic variations between independent sources.  Thus the jump in amplitude between angular separations $\theta = 0.025$ and $>0.075^\circ$ is partially due to the fact that the latter set is necessarily determined using independent sources, with independent intrinsic RMs. The variation in the intrinsic (internal) RMs of independent sources contributes a ``white-noise'' component to the RM variability, which manifests itself as an offset in the amplitude at all points $\theta > 0.075^\circ$. \newline 

From the definition of the structure function in Equation~\ref{SFeq}, it can be easily shown that the difference of two structure functions is equal to the structure function corresponding to the difference of two equivalent power spectra. Furthermore, since the intrinsic RM variations of the sources are independent of any foreground structure, taking the difference between the two structure functions removes the offset. As such, subtracting the RM structure function outside the lobes from the RM structure function behind the lobes (i.e. Figure~\ref{InOutDiffLobe}a $-$ Figure~\ref{InOutDiffLobe}b) gives us a direct measure of the structure function of RM fluctuations due to Centaurus A alone. This is shown in Figure~\ref{InOutDiffLobe}c.  \newline

Figure~\ref{InOutDiffLobe}c shows a significant signal at $\theta > 0.2^\circ$, which implies that the lobes of Centaurus A contribute a measurable RM structure only on scales greater than this.  The signal saturates at $\theta \approx 0.3^\circ$ and, within the present errors, is consistent with a constant amplitude thereafter.  \newline

\begin{figure*}
\centering
\mbox{\subfigure[]{\includegraphics[width=5.8cm]{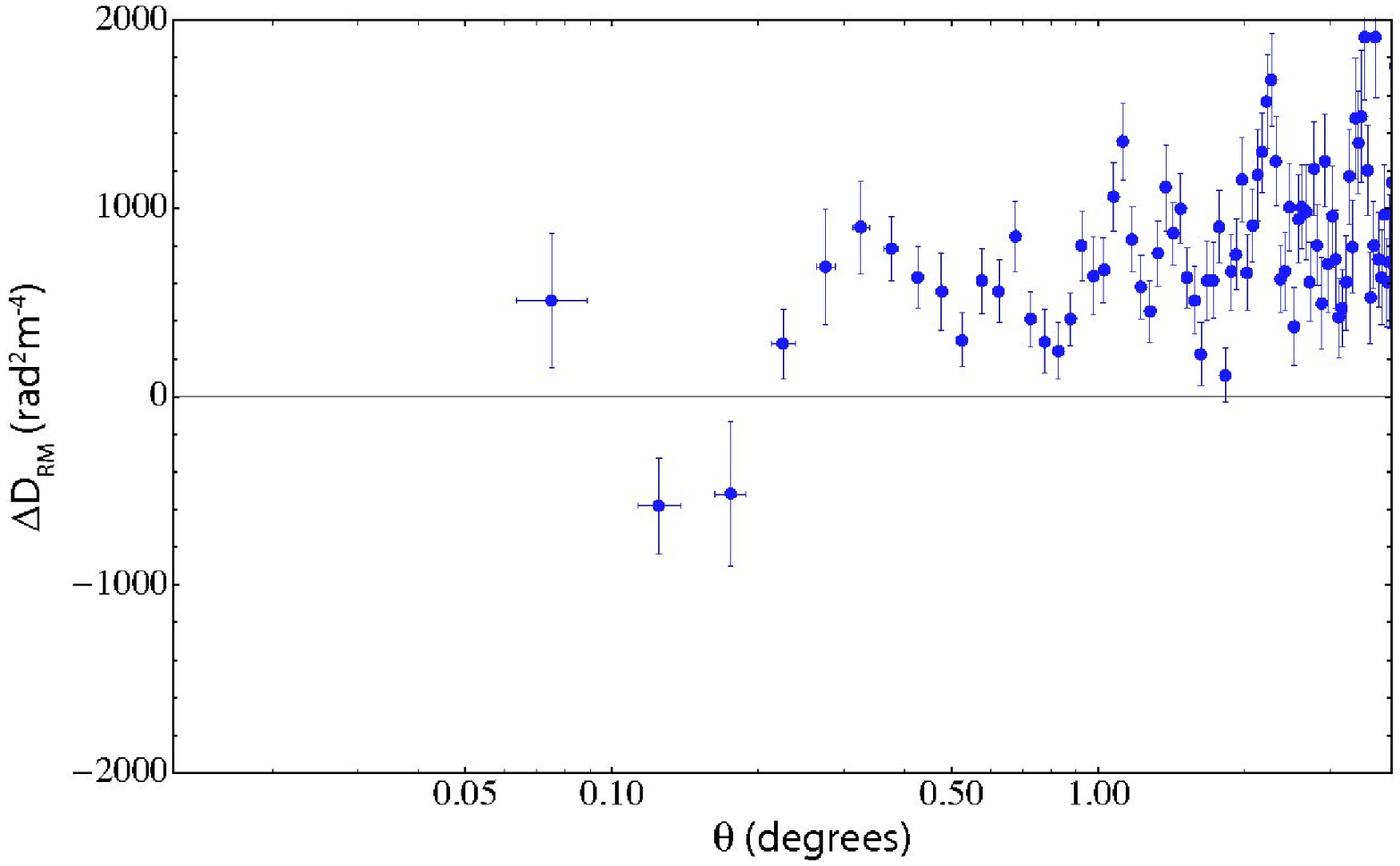}}\quad
\subfigure[]{\includegraphics[width=5.8cm]{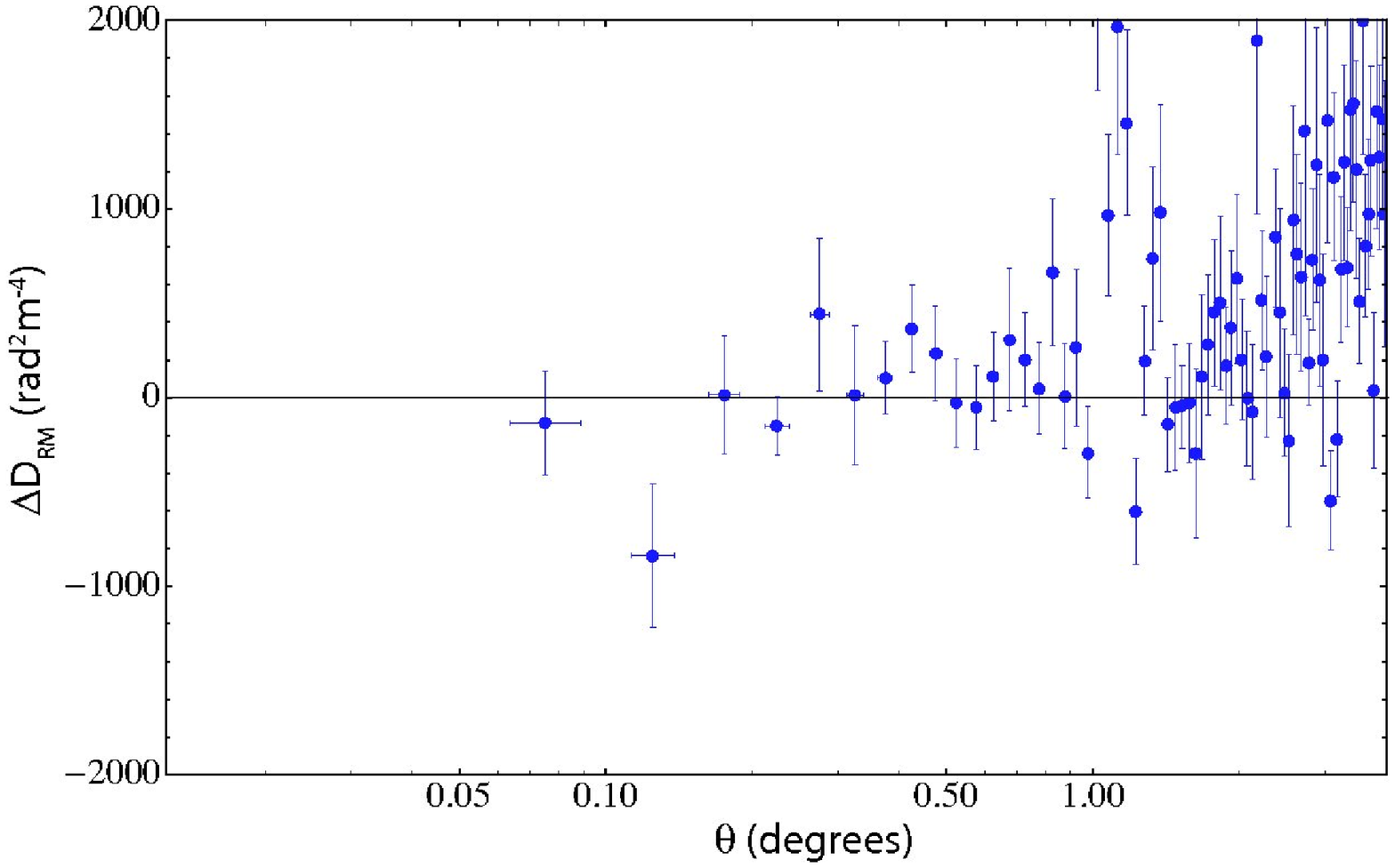}}\quad
\subfigure[]{\includegraphics[width=5.8cm]{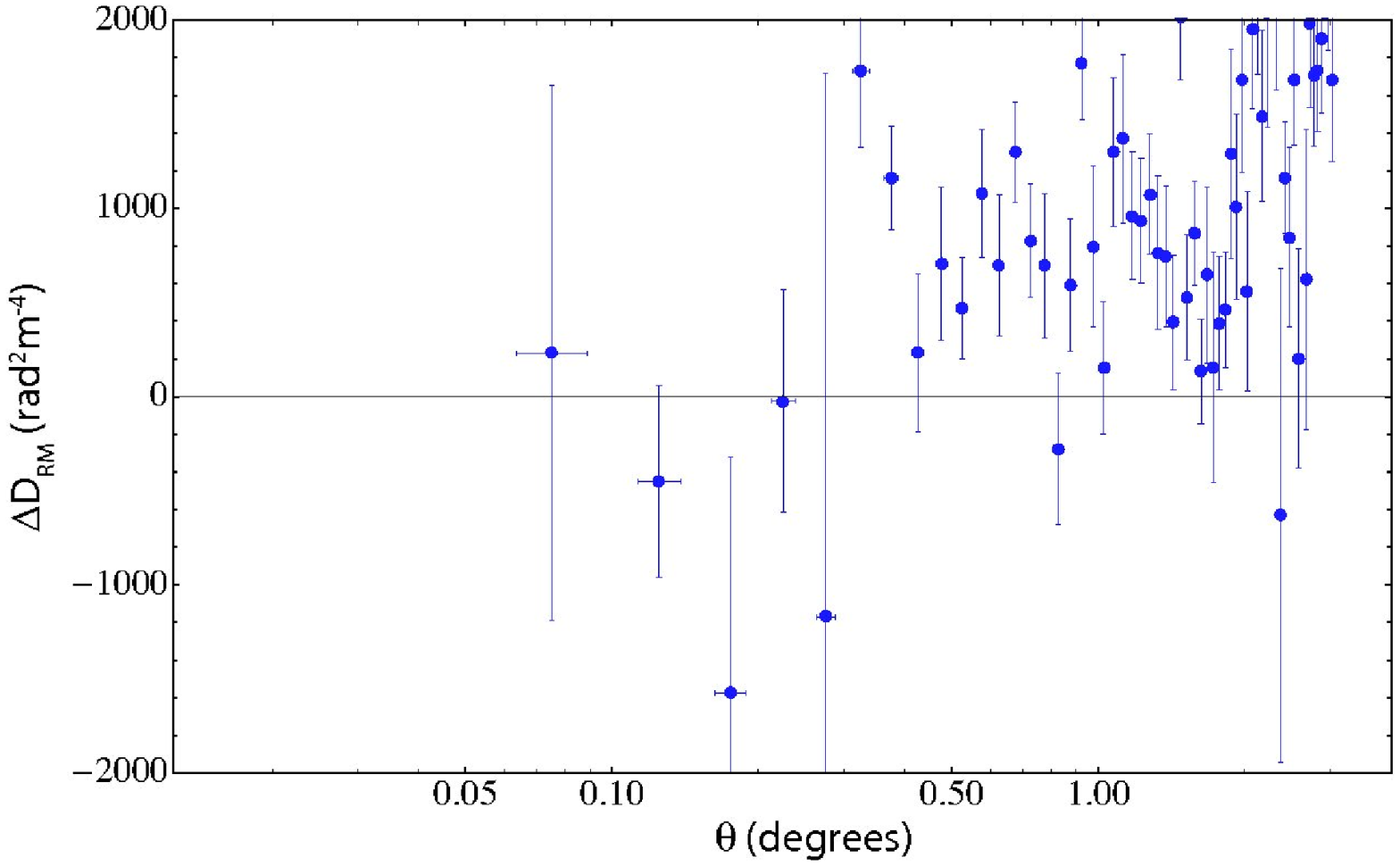}}}
\caption{(a) The RM structure function of the sources that are located north of $\delta=-43$\degree\ subtracted from the RM structure function of all sources that are located south of $\delta=-43$\degree. (b) The structure function of all sources \textit{outside the lobes} that are located north of $\delta=-43$\degree\ subtracted from the RM structure function of all sources \textit{outside the lobes} that are located south of $\delta=-43$\degree. (c) The RM structure function of all sources behind the northern lobe subtracted from the RM structure function of all sources behind the southern lobe.}
\label{NSall}
\end{figure*}

We performed a number of additional tests on the data as a means of establishing the reality of this signal and isolating its origin.  The most obvious cause of a spurious signal is that there is some large scale asymmetry in the spatial rotation measure fluctuations between the northern and southern lobes due to a gradient in the properties of the Milky Way. Figures~\,\ref{NSall}a--c explore this possibility and confirms that the signal in Figure~\ref{InOutDiffLobe}c arises due to RM fluctuations associated with the southern lobe of Centaurus A.  Figure~\,\ref{NSall}a compares the RM structure function of all the polarised sources that are located south of declination $-43$\degree\ to all the polarised sources located north of declination $-43$\degree. There is a clear strong north-south asymmetry present in the data, indicating an excess of RM structure in sources in the south relative to those in the north. Figure\,\ref{NSall}b compares all sources that are \textit{outside the lobes} and located south of $\delta=-43$\degree\ to all sources located \textit{outside the lobes} and north of $\delta=-43$\degree. Figure\,\ref{NSall}b shows no evidence for a north-south asymmetry outside the lobes. Figure\,\ref{NSall}c compares sources located inside the southern lobe to those inside the northern lobe. Here again the asymmetry between the lobes is apparent but the significance is degraded by the smaller number of sources within each lobe. Taken together, Figures~\,\ref{NSall}a-c suggest that the north-south asymmetry is due to variations between the northern and southern lobes of Centaurus A, and not due to the Milky Way. \newline

Note that the source density in each region is accounted for in determining the error bars associated with that structure function and does not otherwise systematically bias the structure function. 

\subsection{Physical interpretation of the lobe RM}\label{proof}
The total and polarised intensity structure of Centaurus A correlates very well with the RM variations of the background sources (Feain et al. 2010, in preparation; \citealp{junkes93}). Such a correlation could not occur if the RM structure is unrelated to Centaurus A itself. For example, \citet{junkes93} investigated the polarised structure of Centaurus A at 6.3\,cm and showed there is a striking difference in the polarised intensity structure between the northern and southern lobes. The polarised emission in northern lobe largely follows the continuum emission uniformly down to the sensitivity limits of the survey. There is little evidence for depolarisation in the northern lobe. The emission in the southern lobe, however, is depolarised \citep{gw71} and chaotic with position angle jumps of up to 90\degree\ in places. Between $\delta=-44$\degree45\arcmin\ and $\delta=-45$\degree45\arcmin\ (J2000 coordinates) the southern lobe becomes highly chaotic and turbulent (see Figure~3b in \citealp{junkes93}). The good spatial coincidence between the depolarization of the southern lobe and that of the RM signal in the background sources on angular scales between $0.2^\circ \lesssim \theta \lesssim 4.0^\circ$ inside the lobes gives us confidence that the depolarising (rotating) medium is intrinsic to the southern radio lobe of Centaurus A. \newline

The amplitude of the RM signal in Figure~\ref{InOutDiffLobe}c is approximately 620~rad$^2$\,m$^{-4}$, implying that the lobes of Centaurus A contribute an RM signal with an rms $\sigma_{RM}\approx 17~$rad\,m$^{-2}$ on an angular scale $\sim$0.3\degree\ (20\,kpc at the distance to Centaurus A). This RM signal could arise either from a thin skin around the lobe or a turbulent medium throughout the lobe. Our data are not sufficient to confidently distinguish between these two possibilities, but we tend to favour the former (thin skin) based on the location of the maximum dispersion of RMs seen in Figure~\ref{rm-distribution} being well aligned with total intensity features along the western boundary of the southern lobe (Feain et al. 2010, in preparation) which look like surface-wave instabilities. We model this signal as follows:\newline

Suppose that the edge of the southern lobe has uniform thermal density, $n^{skin}_{th}$, and a coherent magnetic field, $B_t$, with angle of orientation, $\theta_r$, which changes direction randomly on a spatial scale of $l\sim$\,20\,kpc. Then, following \citet{gaen01}, the dispersion of the RMs produced by unformly polarised emission passing through the edge of the southern lobe is given by:
\begin{equation}
\frac{\sigma_{RM}}{\rm{rad~m}^{-2}} = \frac{810}{\sqrt{3}}\biggl(\frac{n^{skin}_{th}}{\rm{cm}^{-3}}\biggr)\biggl(\frac{B_t}{\mu \rm{G}}\biggr)\sqrt{\biggl(\frac{D}{\rm{kpc}}\biggr)\biggl(\frac{l}{\rm{kpc}}\biggr)}.
\label{eqn:turbulentcom}
\end{equation}
where $D=180$\,kpc is the estimated path length through the edge of the lobe. To estimate $n^{skin}_{th}$ we assume mixing occurs between the lobe magnetic field and the Centaurus intra-group medium due to Kevin-Helmholtz instabilities \citep{bicknel90}.  Based on the ram pressure stripping arguments to explain \textsc{hi} gas depletion in dwarf galaxies in the Centaurus group, \citet{bouch07} infer an intragroup medium density near Centaurus A of $n_{e}\approx10^{-3}$\,cm$^{-3}$; note however that based on X-ray observations, \citet{fei81} estimate a similar density for the interstellar medium of NGC\,5128 itself out to 9\,kpc. Substituting $\sigma_{RM}, D$ and $l$ into Equation~\ref{eqn:turbulentcom} results in $B_t\approx0.8\,n^{-1}_{1}\,\mu$G, where $n^{skin}_{th}=n_1\,10^{-3}$\,cm$^{-3}$. \newline

The possible existence of a magneto-ionic skin around the edges of Centaurus A --- a low power Fanarof-Riley class I (FR\,I) source \citep{fan74} --- with RM fluctuations on scales of 20\,kpc, is not dissimilar to that inferred for skins around the lobes of the powerful FR\,II sources PKS~2104$-$25N and Cygnus~A \citep{bicknel90,cameron88}. In the latter two cases, the RM signals are between 10 and 100 times larger than that seen in Centaurus A, which is not unexpected given both sources are located at or near the centres of clusters where the intracluster density will be higher, possibly by several orders of magnitude.

\section{Conclusion}
\label{conclusion}

In this paper, we have presented the results of a 1.4\,GHz spectropolarimetric imaging survey of 34 deg$^2$ centred on the nearest radio galaxy Centaurus A. A catalogue of the 1005 background, compact radio sources brighter than 3\,mJy~bm$^{-1}$ has been made available in electronic format. \newline

We used Faraday rotation measure (RM) synthesis to measure linear polarised intensities and RMs for each source in our sample. There were 281 sources, out of 1005 in total, with a SNR$\geq$7 in linear polarised intensity and for these we also publish a table of polarised intensities and RMs. The density of RMs published here, 8.3 per deg$^2$, represents the densest RM grid known to date.  Of the 281 RMs, 121 come from sources located behind the radio lobes of Centaurus A; the remaining 160 being from sources located along sightlines outside (in projection) the radio lobes. \newline 

The continuum source counts and the linear polarised intensity (P) source counts (for sources with SNR$\geq$7 in P) are consistent with published source counts giving us confidence that our sample is statistically robust and the emission from the lobes of Centaurus A has not contaminated the global statistics of our sample. There is no measurable difference between the polarised source counts behind the lobes and those outside the lobes. We derived an upper limit on the very small scale turbulence in the lobes of $<10$\,rad\,m$^{-2}$ on scales $\lesssim$180\,pc. This is similar to the limit derived on the volume-averaged RM signal from the lobes. \newline 

We modelled the Milky Way contribution to the RMs by fitting a first order polynomial surface to the 160 RMs along sightlines outside the lobes of Centaurus A and subtracting this surface from all 281 RMs in our sample. The residual RMs behind the lobes have a mean of $-4.5\pm2.8$~rad\,m$^{-2}$ and a standard deviation of 30.4~rad\,m$^{-2}$, compared to a mean of $-1.6\pm2.1$~rad~m$^{-2}$ and standard deviation of 26.6~rad\,m$^{-2}$ for RMs outside the lobes. Assuming a magnetic field strength in the lobes of $1.3\,B_{1}\,\mu$G, we derived an upper limit to the uniform thermal plasma density in the lobes of $\langle n_e \rangle < 5\times10^{-5}\,B_1$ cm$^{-3}$. This is a factor of $2-3$ times smaller than previously published limits and further constrains the prospects for high-energy acceleration processes in the lobes of Centaurus A.\newline

We used a structure function analysis to measure an excess RM dispersion in the southern lobe of Centaurus A of $\sigma_{\textrm RM}\approx 17~$rad\,m$^{-2}$ on angular scales of $\sim0.3$\degree\ associated with the southern lobe of Centuarus A. The RM signal is modelled as possibly arising from a thin skin with a thermal plasma density equivalent to the Centaurus intragroup medium density and a coherent magnetic field that reverses its sign on a spatial scale of 20\,kpc. If the Centuarus intragroup density is of order $10^{-3}$ cm$^{-3}$, the skin magnetic field strength is of order $1~\mu$G. If $n_e$ is an order of magnitude smaller, $B_t$ will be $10\times$ larger. \newline

A more sensitive survey of the southern lobe is now required to verify if the RM signal is associated with the entire southern lobe, or a thin skin (sheath) around the edges, although the latter seems more likely.  In addition, a deep X-ray observation of the edge of the southern lobe would be incredibly revealing in terms of the detailed physics it could probe. The southern lobe of Centaurus A has typically been neglected in favour of investigations of the northern lobe; this is almost certainly because of the very interesting northern middle lobe (NML) region (there has been no southern counterpart to the NML detected). In this paper, we have shown that the edge of the giant southern lobe, rather than any part of the giant northern lobe, is probably associated with fluctuating RMs on scales of 20\,kpc. It is extremely timely now to follow-up this region with a multiwavelength campaign at, in particular, X-ray and deep narrow-band optical wavelength regimes. We are currently completing a detailed analysis of the full polarisation radio continuum images of Centaurus at a resolution of $\approx 40$\arcsec. In this analysis, we will be able to compare the fluctuating RM region with the polarised and total intensity structure of the lobes at a factor of about 5 times the resolution that has been previously possible. If our model of a thin skin is correct, these new images should reveal the signatures of surface waves around the lobes, which would be evidence of mixing between the lobe emission and the Centaurus intragroup medium.\newline

\begin{acknowledgements}
I.J.F. is extremely grateful to Katherine Newton-McGee and Minnie Mao who helped tirelessly with the $\sim1200$ hours of ATCA observations required to complete this project, especially over Christmas to New Year periods in 2006 and 2007. The authors gratefully acknowledge George Heald for providing the RM synthesis CLEAN algorithm. R.D.E. and B.M.G. were recipients of Australian Research Council (ARC) Federation Fellowships (FF0345330 and FF0561298). T.M. acknowledges the support of an ARC Australian Postdoctoral Fellowship (DP0665973). The authors are grateful to the anonymous referee for helpful suggestions that improved this manuscript. The Australia Telescope Compact Array is part of the Australia Telescope which is funded by the Commonwealth of Australia for operation as a National Facility managed by CSIRO.  This research has made use of the NASA/IPAC Extragalactic Database (NED) which is operated by the Jet Propulsion Laboratory, California Institute of Technology, under contract with the National Aeronautics and Space Administration.
\end{acknowledgements}
\bibliography{mnemonic,mnemonic-simple,biblio}

\end{document}